# Genetic drift opposes mutualism during spatial population expansion

Melanie J.I. Müller, Beverly I. Neugeboren, David R. Nelson, and Andrew W. Murray

February 24, 2014

**Abstract**

Mutualistic interactions benefit both partners, promoting coexistence and genetic diversity. Spatial structure can promote cooperation, but spatial expansions may also make it hard for mutualistic partners to stay together, since genetic drift at the expansion front creates regions of low genetic and species diversity. To explore the antagonism between mutualism and genetic drift, we grew cross-feeding strains of the budding yeast S. cerevisiae on agar surfaces as a model for mutualists undergoing spatial expansions. By supplying varying amounts of the exchanged nutrients, we tuned strength and symmetry of the mutualistic interaction. Strong mutualism suppresses genetic demixing during spatial expansions and thereby maintains diversity, but weak or asymmetric mutualism is overwhelmed by genetic drift even when mutualism is still beneficial, slowing growth and reducing diversity. Theoretical modeling using experimentally measured parameters predicts the size of demixed regions and how strong mutualism must be to survive a spatial expansion.

## 1 Introduction

Spatial population expansions are common events in evolutionary history. They range from the growth of microbial biofilms on surfaces [1] to the pre-historic human migration out of Africa [2] and will occur more frequently as climate change forces species to shift their territories [3]. When populations expand, the first individuals to arrive in the new territory are likely to be the ancestors of the later populations in this area. This 'founder effect' produces regions with low genetic diversity because they are occupied by the progeny of a few founders [4]. With few founders, the random sampling of individuals (genetic drift) becomes important. The invasion of different regions by different founders can lead to spatial separation of genotypes ('demixing') [4, 5].

Territorial expansions can have profound effects on the interactions between species or genotypes [6, 7]. For example, the associated demixing can spatially separate cooperators from non-cooperating 'cheaters' [8, 9, 10, 11], in line with the common view that spatial structure in general enhances cooperation. In contrast, spatial demixing may have a detrimental effect on mutualistic interactions (beneficial for both partners). Mutualism selects for coexistence ('mixing') of the two partners [12], as was recently shown for a microbial mutualism in a spatial setting [13], and theory argues that the demixing caused by spatial expansion can extinguish mutualism [14, 15]. Mutualism imposes constraints on spatial expansions: Obligate mutualists must invade new territory together, and facultative mutualists invade faster when mixed.

Despite these constraints, major events in evolutionary history involve spatial expansions of mutualists. The invasion of land by plants may have taken advantage of the mutualistic association with fungi [16], and flowering plants spread with their pollen-dispersing insects [17]. More recently the invasion of pine trees in the Southern hemisphere required mycorrhizal fungal symbionts [18], and legumes can only grow in new areas with their mutualist nitrogen-fixing rhizobacteria [19]. Microbes in biofilms often exhibit cooperative interactions [20], such as interspecies cooperation during tooth colonization [21]. A common microbial mutualism is cross-feeding, i.e. the exchange of nutrients between species [22, 23, 24, 25, 26, 27].

Here, we use the growth of two cross-feeding strains of the budding yeast *Saccharomyces cerevisiae* on agar surfaces to study the antagonism between genetic drift and mutualism during spatial expansions. The strains exchange amino acids, allowing us to control the mutualism's strength by varying the amino acid concentrations in the medium. The strains demix under non-mutualistic conditions, but, for obligate mutualism, expand in a more mixed pattern whose characteristics we explain with a model of the nutrient exchange dynamics. When mutualism is facultative or highly asymmetric, genetic drift dominates, leading to demixing even when mixing would be beneficial. We quantitatively understand this transition using a generalized stochastic Fisher equation.

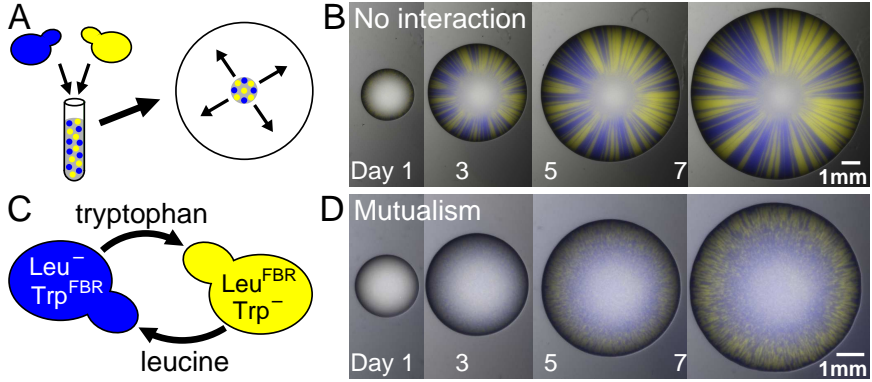

Figure 1: **A**) Spatial expansion assay. Two fluorescently labeled *S. cerevisiae* strains, depicted as blue and yellow, are mixed in liquid and pipetted as a circular drop onto an agar surface. When the colony expands, the ensuing spatial pattern can be monitored by fluorescence microscopy. **B**) Successive images of the expansion of two non-interacting yeast strains show the formation of distinctive blue and yellow sectors. **C**) Two cross-feeding yeast strains as a model for mutualism. The yellow strain $\text{Leu}^{\text{FBR}}\,\text{Trp}^-$ produces leucine but not tryptophan, while the blue strain $\text{Leu}^-\,\text{Trp}^{\text{FBR}}$ produces tryptophan but not leucine. To grow on medium lacking both amino acids, the strains must cross-feed each other. The strains are feedback-resistant (FBR) in the production of leucine or tryptophan, leading to increased production and therefore secretion of these amino acids. **D**) These mutualistic strains form small, intertwined patches during spatial expansion, see also Fig. 2C.

## 2 Results

To study spatial expansions, Hallatschek *et al.* pioneered a simple microbial expansion assay [5]. Two yeast strains labeled with two different fluorescent proteins, depicted as yellow and blue in Fig. 1A, are mixed and inoculated as a circular drop (the 'homeland') on an agar surface. The colony grows radially outwards on the surface as cell division pushes cells forward (yeast has no active motility). The cells deplete the nutrients in the agar immediately below the colony, and then grow solely on nutrients diffusing towards the colony from the surrounding agar, restricting growth to a small 'active layer' extending only $40\,\mu$m back from from the colony boundary [28, 29]. The small number of cells involved in local colony propagation leads to a high local fixation probability for blue or yellow cells [5, 30] (Fig. 1B). Colony expansion reduces diversity: a front that migrates from a well-mixed homeland produces sectors that are fixed for yellow or blue cells.

### 2.1 Strong mutualism inhibits demixing

To study mutualism, we genetically engineered the yeast strains shown in Fig. 1C. These strains cross-feed each other two amino acids, leucine (leu) and tryptophan (trp). To enhance cross-feeding, we used previously characterized feedback-resistant (FBR) mutations [31, 32] that increase amino acid production by inactivating the feedback inhibition that normally regulates amino acid production. The strain $\text{Leu}^{\text{FBR}}\,\text{Trp}^-$, depicted as yellow, overproduces leucine ($\text{Leu}^{\text{FBR}}$) and leaks it into the medium, but cannot produce tryptophan ($\text{Trp}^-$). Its partner strain $\text{Leu}^-\,\text{Trp}^{\text{FBR}}$, depicted as blue, overproduces and leaks tryptophan ($\text{Trp}^{\text{FBR}}$) but cannot produce leucine ($\text{Leu}^-$). Because growth requires both leucine and tryptophan, neither strain can grow on medium lacking both amino acids.

When mixed together, the two cross-feeding strains grow robustly on medium without leucine and tryptophan, forming interdigitated patches (Fig. 1D). Since each strain needs the amino acid from its partner strain to proliferate, they cannot demix into the large separated sectors seen for non-interacting strains, but must stay in close proximity. However, some segregation still occurs: the mutualists make visible yellow and blue patches (Fig. 1D), but these patches are much smaller than the sectors of non-interacting strains (Fig. 1B).

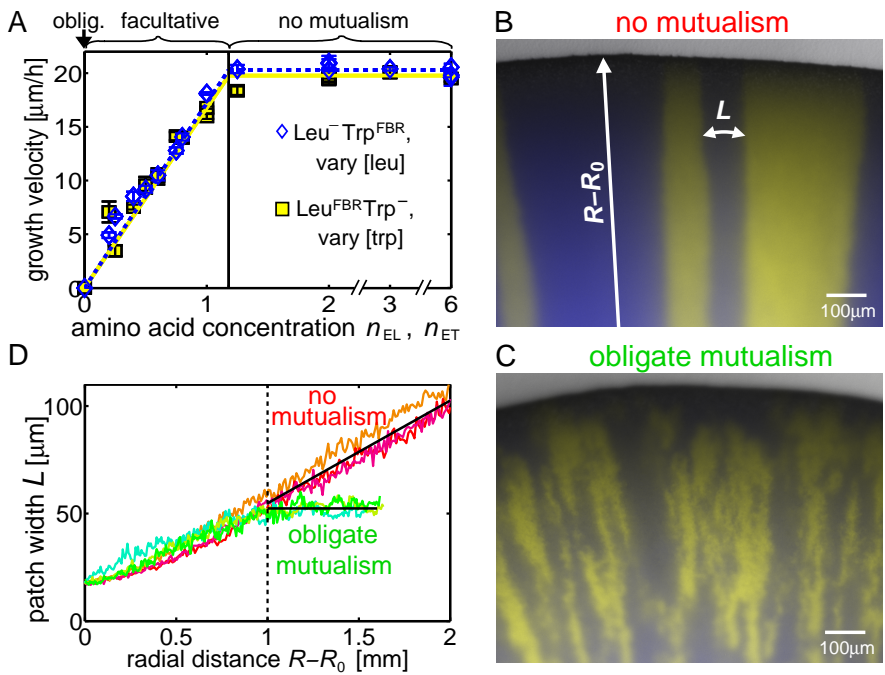

Figure 2: **A)** The radial growth velocities of single-strain colonies of $\text{Leu}^-\text{Trp}^{\text{FBR}}$ (blue diamonds) and of $\text{Leu}^{\text{FBR}}\text{Trp}^-$ (yellow squares) increase linearly with the leucine concentration $n_{\text{EL}} = [\text{leu}]/[\text{leu}]_c$ and the tryptophan concentration $n_{\text{ET}} = [\text{trp}]/[\text{trp}]_c$ in the medium, respectively, until they saturate at a plateau. This behavior is indicated by the corresponding blue dotted and yellow solid lines, which are piecewise linear fits. Concentrations are scaled with the factors $[\text{leu}]_c = 762\,\mu\text{M}$ and $[\text{trp}]_c = 98\,\mu\text{M}$, to make the crossover between the linear and the plateau regime occur at the same rescaled concentrations (vertical line). For concentrations above this crossover value, mutualism is irrelevant ('*no mutualism*'). Mutualism is *facultative* for lower and *obligate* for zero concentrations. **B)** For no mutualism, colonies exhibit demixing into large sectors. **C)** For obligate mutualism, colonies form much smaller, intertwined patches. **D)** Average width $L$ (parallel to the front) of yellow and blue patches as function of the radial distance $R-R_0$ from the homeland (perpendicular to the front) for three replicate colonies under conditions of no (three independent, reddish lines) and obligate (three independent, greenish lines) mutualism. For the first mm, $L$ increases due to genetic demixing and sector boundary diffusion. Afterwards, obligate mutualism limits the patch width to $L = 52\,\mu\text{m}$ (black horizontal line), while the sector width increases linearly with the radius for no mutualism (black inclined line).

Genetic drift and mutualism are opposing forces: mutualism mixes and drift demixes. To probe this antagonism, we change the strength of mutualism by varying the levels of leucine and tryptophan in the medium. We quantify this effect by measuring the expansion velocities of single-strain colonies for different concentrations of the amino acid they need, e.g. leucine for $\text{Leu}^-\text{Trp}^{\text{FBR}}$(Fig. 2A). As expected, strains cannot grow without their required amino acid; as the amino acid concentration increases, the velocity increases approximately linearly and then plateaus. The surprising linearity below the plateau suggests that cells alter the number and/or the affinity of amino acid transporters in response to the external amino acid concentration. Appropriately scaling the amino acid concentrations makes velocity plots of the two strains overlap almost exactly. The leucine scaling factor, $762\,\mu\text{M}$, is 8 times larger than the tryptophan scaling factor, $98\,\mu\text{M}$, consistent with yeast proteins containing ∼10 times as much leucine as tryptophan [33]. The small deviation of the vertical crossover line from a value of 1 in Fig. 2A is due to amino acid loss into the medium as explained in SI Sec. S2.

The effects of amino acid concentration on growth define three growth regimes. On medium without leucine and tryptophan, the two strains can grow together but not alone and thus form a pair of *obligate mutualists*. For increased amounts of leucine and tryptophan, mutualism becomes *facultative*, since the strains get amino acids from the medium as well as from their partner (that the partner's presence still leads to faster growth in this regime is shown below in Fig. 4F). Above a critical concentration (the onset of the plateau regime of Fig. 2A), leucine and tryptophan are no longer growth limiting, and amino acids leaked by the cells should not matter. We therefore expect cells to demix into well-defined sectors like the non-interacting cells in Fig. 1B. This is indeed the case (Fig. 2B), defining this as the *no mutualism* regime.

## 2.2 Patterns for obligate and no mutualism

We first studied the extreme cases of obligate and no mutualism with their striking difference in expansion patterns (Fig. 2B,C). To quantify this difference, we used image analysis to determine how the average width $L$ (parallel to the front) of patches of a single color changes as expansion progresses for increasing radial distances $R - R_0$ from a homeland of radius $R_0$. As shown in Fig. 2D, for both obligate and no mutualism the patch width initially increases as unicolored patches form by local fixation events due to genetic drift. In addition, patch boundaries diffuse and create larger and larger patches when they collide [5, 34]. For radii $R$ larger than twice the homeland radius $R_0 \approx 1$ mm, the no-mutualism sector width increases linearly with the radius because the sector width increases with the growing colony's circumference, preventing further sector boundary collisions so that the sector number stays constant [34].

For obligate mutualism the patch width plateaus at $L = 52 \pm 1\mu$m, even for large radii. Mutualism requires physical proximity of the interacting partners. For example, cells in a patch of leucine-requiring cells get leucine by diffusion from neighboring leucine-producing patches. The patch cannot get too big. If it did, the cells at its center would starve, because the more peripheral cells would take up all the leucine diffusing from the neighboring patches of leucine producers.

We set out to combine theory and experimentally measured parameters to understand the patch width and other characteristics of the expansion pattern from the nutrient exchange dynamics. The dynamics of the leucine concentration $[\text{leu}](x, t)$ in a reference frame that moves with the front is

$$\frac{\partial [\text{leu}]}{\partial t} = D_\text{a} \frac{\partial^2 [\text{leu}]}{\partial x^2} - d\left([\text{leu}] - [\text{leu}]_\text{E}\right) + r_\text{L} c_\text{T} - K_\text{L}([\text{leu}]) c_\text{L}. \quad (1)$$

The first term describes leucine diffusion, with diffusion constant $D_\text{a}$, along the coordinate $x$ parallel to the front. The second, chemostat-like term is an effective description of leucine diffusion perpendicular to the front. The gradient between the leucine concentration $[\text{leu}]$ at the colony boundary and the leucine concentration $[\text{leu}]_\text{E}$ in the medium far away from the colony generates a diffusive flux away from the colony with the diffusive rate $d$ (see SI Sec. S5 for a detailed derivation). The third term describes secretion of leucine by leucine-producing cells, whose concentration is $c_\text{T}$, with rate $r_\text{L}$ per cell, while the last term describes leucine uptake by leucine-requiring cells, whose concentration is $c_\text{L}$, with a concentration-dependent rate $K_\text{L}([\text{leu}])$. To maintain constant intracellular amino acid concentrations during steady-state growth, the leucine uptake rate has to be proportional to the growth rate. The dependence of colony growth velocity on external amino acid concentration (Fig. 2A) thus motivates a constant uptake rate $K_\text{L}([\text{leu}]) = k_\text{L}$ for concentrations larger than the crossover value $[\text{leu}]_\text{c}$, and linear $K_\text{L}([\text{leu}]) = k_\text{L}[\text{leu}]/[\text{leu}]_\text{c}$ for smaller concentrations. Such a limiting behavior for small and large concentrations is also expected for other functional forms of the uptake rate such as Michaelis-Menten kinetics.

We write Equation [1] in a non-dimensionalized form for the rescaled concentration $n_\text{L} \equiv [\text{leu}]/[\text{leu}]_\text{c}$ by expressing time in units of the inverse diffusive rate $d$ and space in units of the diffusion length scale $\sqrt{D_\text{a}/d}$:

$$\frac{\partial n_\text{L}}{\partial \tilde{t}} = \frac{\partial^2 n_\text{L}}{\partial \tilde{x}^2} - (n_\text{L} - n_\text{EL}) + \rho_\text{L}(1-f) - \kappa_\text{L} T(n_\text{L}) f. \quad (2)$$

Here, we have written the cell concentrations $c_\text{L}$ and $c_\text{T}$ in terms of the fraction $f \equiv c_\text{L}/(c_\text{L} + c_\text{T})$ of leucine-requiring cells, where $c = c_\text{L} + c_\text{T}$ is the constant surface carrying capacity (yeast colonies grow to a finite height). We have defined the function $T(n_\text{L}) = n_\text{L}$ for $n_\text{L} \leq 1$ and $= 1$ for $n_\text{L} > 1$, and the dimensionless secretion and uptake parameters

$$\rho_{\mathrm{L}} \equiv \frac{c}{d}\frac{r_{\mathrm{L}}}{[\mathrm{leu}]_{\mathrm{c}}} \quad \text{and} \quad \kappa_{\mathrm{L}} \equiv \frac{c}{d}\frac{k_{\mathrm{L}}}{[\mathrm{leu}]_{\mathrm{c}}}. \tag{3}$$

Similar equations hold for tryptophan with the rescaled secretion and uptake rates $\rho_{\mathrm{T}} \equiv (c/d)(r_{\mathrm{T}}/[\mathrm{trp}]_{\mathrm{c}})$ and $\kappa_{\mathrm{T}} \equiv (c/d)(k_{\mathrm{T}}/[\mathrm{trp}]_{\mathrm{c}})$. We find that these parameters are equal for our two strains, $\rho_{\mathrm{L}} = \rho_{\mathrm{T}} \equiv \rho$ and $\kappa_{\mathrm{L}} = \kappa_{\mathrm{T}} \equiv \kappa$ (SI Sec. S1.4). Thus, the mutualism described by Equation [2] and the corresponding tryptophan equation is a symmetric interaction, unless asymmetries are introduced via the external concentrations $n_{\mathrm{EL}}$ and $n_{\mathrm{ET}}$. This symmetry is not trivial (below we will see that it does not hold away from steady-state growth), but presumably not accidental. Since yeast cells contain an order of magnitude more leucine than tryptophan, we expect uptake and secretion to be an order of magnitude larger for leucine than for tryptophan. Indeed, the equalities $\rho_{\mathrm{L}} = \rho_{\mathrm{T}}$ and $\kappa_{\mathrm{L}} = \kappa_{\mathrm{T}}$ follow from the scaling relations $k_{\mathrm{L}} = 8\,k_{\mathrm{T}}$, $r_{\mathrm{L}} = 8\,r_{\mathrm{T}}$, and $[\mathrm{leu}]_{\mathrm{c}} = 8\,[\mathrm{trp}]_{\mathrm{c}}$ in our system (SI Sec. S1.4).

According to the symmetry for obligate mutualists ($n_{\mathrm{EL}} = n_{\mathrm{ET}} = 0$), blue and yellow patches should have the same average widths, which we observe (Fig. 2C and SI Fig. S4A). Although mutualist patches are less clearly defined than demixed non-mutualist sectors, we first assume, for simplicity, that a blue patch contains only leucine consumers ($f = 1$). Leucine diffuses into such a patch from neighboring yellow leucine producer patches, but is lost due to uptake and diffusion away from the colony with the combined rate $c\,k_{\mathrm{L}}/[\mathrm{leu}]_{\mathrm{c}} + d = d\kappa + d$. On the loss time scale $t_{\mathrm{loss}}$, which equals the inverse loss rate, leucine diffuses a distance $\sqrt{D_{\mathrm{a}} t_{\mathrm{loss}}} = \sqrt{D_{\mathrm{a}}/(d\kappa + d)}$ into the consumer patch. More rigorously, according to Equation [2] the leucine concentration within the patch decreases exponentially with the distance from the neighboring leucine-producing patches on the length scale

$$l_{\mathrm{a}} = \sqrt{\frac{D_{\mathrm{a}}/d}{\kappa + 1}} \tag{4}$$

This gradient would lead to the patch growing more slowly in its middle than at its boundary, which would cause an unstable, undulating front rather than the smooth front seen in Fig. 2C. Thus, patches must be small compared to the length scale $l_{\mathrm{a}}$ that describes the fall of nutrient concentration within a patch. Indeed, for our system $l_{\mathrm{a}} \approx 700\,\mu\mathrm{m}$, an order of magnitude larger than the mutualistic patch width $L \approx 50\,\mu\mathrm{m}$.

Patch boundaries wander perpendicular to the expansion direction due to the jostling of cell division [34]. Boundary diffusion can 'smooth out' velocity differentials caused by amino acid diffusion provided that both processes happen on comparable length and time scales. A quantitative calculation (SI Sec. S4) shows that this is the case for an average patch width

$$L = 2\sqrt{l_{\mathrm{a}} l_{\mathrm{b}}}, \tag{5}$$

which is twice the geometric mean of the nutrient diffusion length scale $l_{\mathrm{a}}$ and the patch boundary diffusion length $l_{\mathrm{b}} = 2D_{\mathrm{s}}/b$. Since the cellular diffusion constant $D_{\mathrm{s}}$ (a few $\mu\mathrm{m}^2/\mathrm{h}$) is much smaller than the amino acid diffusion constant $D_{\mathrm{a}}$ (a few $\mathrm{mm}^2/\mathrm{h}$), the length scale for the diffusion of the boundary between the two cell types in the active layer of size $b$ is only $l_{\mathrm{b}} \approx 1\,\mu\mathrm{m}$, compared to $l_{\mathrm{a}} \approx 700\,\mu\mathrm{m}$. From Equation [5], we estimate $L \approx 50\,\mu\mathrm{m}$, which agrees well with the value observed in Fig. 2.

Due to patch boundary diffusion, patches have ragged boundaries and are not completely demixed. Patches that look yellow contain blue cells and vice versa, as judged by comparing their fluorescence intensities with those of fully demixed blue and yellow sectors, see below. A second argument comes from the patch width of obligate mutualists remaining constant as the colony grows (Fig. 2D). Since the circumference increases during radial expansion, the number of patches at the circumference must increase to maintain a constant patch width. Indeed, we see new yellow patches emerge from within blue patches (Fig. 2C), which is only possible if the blue patch contain some yellow cells. Similarly, blue patches emerge from within yellow patches in Fig. 2C.

Incomplete demixing is due to frequency-dependent selection that promotes stable coexistence of two interacting strains [12, 14]. In our system, the selection coefficient

$$s(f) = \frac{V_{\mathrm{L}}(f) - V_{\mathrm{T}}(f)}{fV_{\mathrm{L}}(f) + (1-f)V_{\mathrm{T}}(f)}, \tag{6}$$

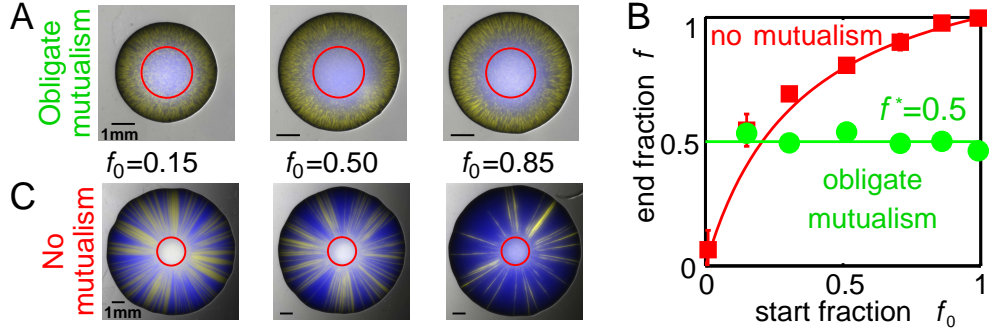

Figure 3: **A)** Obligate mutualists grow into a characteristic patchy pattern independent of the start fraction $f_0$ in the homeland (red circle). **B)** The final fraction at the colony boundary equals $f = f^* = 0.5$ for obligate mutualism (green circles), but depends on the start fraction $f_0$ for no mutualism (red squares). Solid lines are solutions to selection dynamics with selection coefficient $s(f)$ of Equation [6] for obligate and $s = 0.02$ for no mutualism (blue cells are 2% more fit than yellow cells). **C)** Non-mutualistic colonies expand from the homeland (red circle) into demixed sectors whose number and width depends on the start fraction.

is frequency-dependent because the growth velocities $V_L$ and $V_T$ of the two strains depend on the amino acid concentrations $n_L$ and $n_T$, which in turn depend on the cellular fraction or 'allele frequency' $f$ via Equation [2] and the corresponding tryptophan equation. Selection drives the system towards a stable fraction $f^*$ with equal velocities, $V_L(f^*) = V_T(f^*)$. Since our mutualistic interaction is symmetric, we predict $f^* = 0.5$.

This prediction was confirmed experimentally: obligate mutualists inoculated with different starting fractions $f_0$ expand into the same characteristic pattern with average fraction $f^* = 0.5$, independent of $f_0$ (Fig. 3A,B). The transient dynamics towards steady-state growth depends on the start fraction: colonies with $f_0 = 0.15$ are smaller because they take longer to start growing (SI Sec. S1.3), and colonies with $f_0 = 0.01$ do not grow at all. While our simple model for steady-state growth captures the time scale of this transient, it does not capture these asymmetric features (SI Sec. S3.3).

For no mutualism, selection is not frequency-dependent, and the colony boundary fraction $f$ depends on the start fraction $f_0$ (Fig. 3B,C). The fraction $f$ of blue cells increases during expansion because the blue tryptophan producers have a 2% fitness advantage over the yellow leucine producers under these conditions (Fig. 2A and SI Sec. S1.2), presumably because overproduction of tryptophan, a rare amino acid, is less costly than overproduction of the more abundant leucine.

## 2.3 Genetic drift can overcome facultative mutualism

That obligate mutualists remain (partially) mixed during spatial expansion is not completely unexpected since they *must* remain together to grow. We next study facultative mutualists, which can invade new territory on their own. Motivated by Fig. 2A, we decrease the mutualistic strength by increasing the amino acid concentrations in the medium. Fig. 4A shows the resulting colonies. For low leucine and tryptophan concentrations, expansion produces mixed patches with ragged boundaries, whereas large, well-separated sectors with smooth boundaries appear for high concentrations. When one amino acid is more abundant than the other, the strain requiring this amino acid has an advantage and dominates the colony. This behavior can also be seen in the colony boundary fraction $f$ of blue cells (Fig. 4B). In summary, weak or asymmetric mutualism (high or asymmetric amino acid concentrations) leads to demixing, and strong mutualism (low concentrations) leads to mixing.

To probe the mutualism-drift antagonism, we focus on amino acid concentrations that retain the symmetry of our interaction, i.e. colonies whose boundary fractions $f$ equal the inoculation fraction $f_0 = 0.5$. These colonies (outlined in white in Fig. 4A) are slightly off the diagonal $n_{EL} = n_{ET}$, presumably because of the slight fitness advantage of blue cells and the approximate nature of the linear fit in Fig. 2A.

To study the degree of mixing, we consider the local fraction $f$ of blue cells at the colony boundary. In the demixed case, a histogram of $f$ should have peaks at $f = 1$ (from blue sectors) and $f = 0$ (from yellow sectors), with only few $0 < f < 1$ values (at sector boundaries). In contrast, more mixed mutualistic patches should exhibit a broad peak around $f = 0.5$. The yellow fluorescence intensity, a measure of the amount of yellow cells, displays this behavior (Fig. 4C): for low external amino acid concentrations, the histograms have a single peak, while they are clearly bimodal for higher concentrations. More quantitatively, as mutualism becomes weaker with increasing amino acid concentration $n_E \equiv (n_{EL} + n_{ET})/2$, the probability of seeing one peak in histograms of random subsets of the fluorescence intensity data drops sharply from 1 to 0 as $n_E$ increases above 0.25 (Fig. 4D). Correspondingly, the patch width $L$ increases with $n_E$ until it plateaus for $n_E \gtrsim 0.25$ (Fig. 4E). We conclude that genetic drift overpowers mutualism for amino acid concentrations $n_E \gtrsim 0.25$, even though mutualism and complete mixing would still be beneficial, since $n_E = 0.25$ supports only 25% of the maximal growth rate of isolated, well-fed strains (Fig. 2A and Fig. 4F).

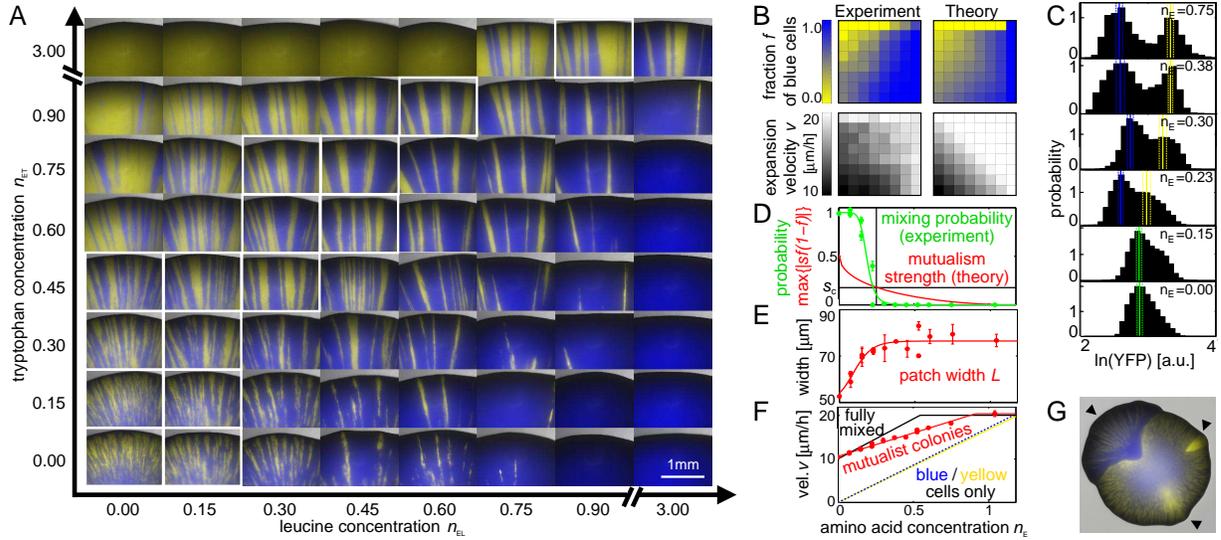

Figure 4: Tuning strength and symmetry of mutualism.
**A)** Images of colony boundaries for different external leucine and tryptophan concentrations $n_{EL}$ and $n_{ET}$ in the medium. For low concentrations (lower left corner), colonies display a patchy pattern with ragged boundaries, while high external amino acid concentrations (upper right corner) lead to demixing into clear sectors with smooth boundaries. If leucine is abundant but tryptophan is not (lower right corner), the blue leucine-requiring strain wins, whereas abundant tryptophan and low leucine (upper left corner) favors the yellow tryptophan-requiring strain. Colonies with boundary fractions $f$ within 10% of the inoculation fraction $f_0 = 0.5$ are outlined by white squares and further investigated in (C-F).
**B)** Average colony boundary fraction $f$ of blue cells (top) and expansion velocity $v$ (bottom), as measured experimentally (left) and predicted by a model for well-mixed growth (right). The horizontal and vertical axes cover the same amino acid concentrations as in (A).
**C)** Characterization of demixing along the 'diagonal' (white-square images in (A)). The yellow fluorescence intensity histograms display a single broad peak for small $n_E = (n_{EL} + n_{ET})/2$, and two peaks for large $n_E$. Vertical lines indicate the mode locations (solid lines) and their standard deviations (dashed lines) for histograms from random data subsets.
**D)** The probability of finding 1 peak (green points, green line to guide the eye) in fluorescence histograms of random data subsets decreases with increasing $n_E$, exhibiting a sharp drop near $n_E = 0.25$ (vertical black line). The theoretically predicted strength of mutualism, estimated as the maximum of the selection term $|s(f)f(1-f)|$ of Equation [6] (red line), becomes comparable to the strength of genetic drift, $s_c = D_g^2/D_s$ (horizontal black line), at the concentration $n_E = 0.25$ of crossover from mixing to demixing.
**E)** The average width $L$ (red points, red line to guide the eye) of patches of a single color at distance $R - R_0 = 1.5$ mm from the homeland correspondingly increases with $n_E$ until it plateaus for $n_E \gtrsim 0.25$.
**F)** The velocity of mutualistic colonies increases with $n_E$ (red data points), but not as fast as predicted by the mutualistic benefits (black solid line). However, it is always larger than the single strain growth velocities (yellow solid and blue dotted lines replotted from Fig. 2A).
**G)** A colony of obligate mutualists with 3 mutant sectors indicated by black arrowheads. The average is one mutant sector per colony.

To understand how varying the mutualism's strength controls the antagonism between mutualism and genetic drift, we write down a model for the dynamics of the cellular fraction $f(x, \tau)$ of blue cells [35, 36, 14]:

$$\frac{\partial f}{\partial \tau} = s(f) f(1-f) + D_s \frac{\partial^2 f}{\partial x^2} + \sqrt{D_g f(1-f)} \, \Gamma(x, \tau), \qquad (7)$$

where $\tau$ is time measured in generations and $x$ is the coordinate along the front. The first term incorporates mutualism with the selection coefficient $s(f)$ of Equation [6]. The second term describes cellular diffusion due to the jostling of cell division with diffusion constant $D_s$. The last term describes genetic drift, where $\Gamma(x, \tau)$ is an Itō delta-correlated Gaussian noise. The genetic diffusion constant $D_g$ characterizes the noise strength and is expected to be inversely proportional to the effective population density at the front [35].

The selection term favors mixing as long as the amino acid concentrations are below the crossover values in Fig. 2A. In this regime, amino acids secreted by the cells increase growth velocities, so that mutualists benefit from remaining mixed at an optimal fraction $f^*$. Considering only this term gives reasonable predictions for the average cell fraction $f$ (top panels of Fig. 4B) as $f$ depends mainly on the overall amino acid balance. However, it overestimates the growth velocities (compare the bottom panels in Fig. 4B, and the black solid line with the red data points in Fig. 4F). This discrepancy arises because genetic drift during spatial expansion leads to local deviations from the optimal fraction $f^*$, thus slowing colony growth and producing blue and yellow patches instead of a homogeneous mix at fraction $f^*$.

A detailed analysis of Equation [7], assuming locally flat fronts, shows that mutualism loses to genetic drift when the mutualistic strength falls below a critical value $s_c = D_g^2/D_s$ that characterizes the strength of local demixing [14]. Using our independently determined experimental parameters, the crossover concentration is predicted to be $n_E = 0.25$, which is consistent with our experimental observations (Fig. 4C-D). For asymmetric mutualism, e.g. due to more leucine than tryptophan in the medium, local fixation of leucine consumers ($f = 1$) becomes more likely because the mutualistic fixed point $f^*$ is closer to $f = 1$, and because the selective barrier to fixation at $f = 1$ is lower (SI Sec. S3 and Ref. [37]).

Obligate mutualists grow in a characteristic pattern determined by mutualism parameters such as nutrient uptake and secretion rates. On evolutionary time scales, mutations can change these properties and thus the pattern. To our surprise, mutant sectors arise even in our ∼50 generation expansions (Fig. 4G). These mutants presumably change the mutualistic interaction since they expand with patterns different from the ancestors, and since most sectors have a different shape than expected for frequency-independent selection coefficients [38]. We plan to investigate these mutants further in the future.

## 3 Conclusions

During spatial population expansions, mutualism and genetic drift act as antagonistic evolutionary forces. Mutualists benefit from coexistence ('mixing'). Due to this constraint, genetic drift at the expansion front can impede or even destroy mutualism by creating regions that are colonized predominantly or exclusively by one of the partners ('demixing'). We experimentally and theoretically investigated this antagonism during the spatial expansion of two cross-feeding yeast strains. We find that strong mutualism suppresses demixing, but weaker mutualism is overpowered by genetic drift even though the resultant demixing makes the population grow more slowly than it would if it remained fully mixed. A critical mutualistic selection strength is required for mutualists to 'survive' spatial expansions. Our results are quantitatively explained by a model that incorporates mutualistic frequency-dependent selection due to nutrient exchange as well as the diffusional drift of cells due to cell division.

Spatial demixing is particularly pronounced in our experimental system because yeast lack motility and only 'disperse' offspring locally by cell division. For other organisms, movement of individuals and offspring dispersal provide additional mixing. But movement and dispersal are usually spatially restricted, and genetic demixing will occur if expansion into new territory is sufficiently fast compared to migration and dispersal within occupied areas [4, 34]. Spatial sectoring has been observed for mutualists in nature [39, 40].

The detrimental effect of spatial expansion on mutualism contrasts with the notion that spatial structure in general [41, 25] and spatial expansion in particular [8, 9, 10, 11] promote cooperation. In these studies, cooperators benefit from demixing, which separates them from non-cooperating 'cheaters'. In contrast, mutualists profit from coexistence rather than separation and are impeded by expansions. The effect of spatial expansion, similar to spatial structure for stationary populations [42, 43], therefore depends on whether the cooperative interaction selects for a mixture of genotypes. If a mixture of genotypes grows fastest, spatial expansion impedes proliferation by separating these genotypes.

The difficulty of successfully expanding into new territories may contribute to mutualism breakdown [7], and explain the rareness in nature of mutualisms that form exclusively between two species [44]. Our observations suggest that mutualists can only expand their range together if mutualistic benefits are very strong, or if they can ensure coordinated dispersal of the mutualistic partners. Strong benefits presumably allowed the spread of flowering plants and their pollinators [17], the invasion of land by plants with mutualistic fungi [16], and underlie many current plant-microbe mutualisms [18, 19]. Other mutualisms exhibit permanent physical linkage, most notably eukaryotic cells and their mitochondria and chloroplasts, endosymbionts, and lichens. Indirect physical linkage can preserve mutualisms, such as agricultural ants transporting their fungal crops to new nests [45]. In summary, the requirement to 'survive' territorial expansions may have played an important role in the evolution of many mutualisms.

## 4 Materials and methods

**Strains** The haploid, asexually reproducing, *S. cerevisiae* strains Leu$^{\text{FBR}}$ Trp$^-$ and Leu$^-$ Trp$^{\text{FBR}}$ are derived from W303, can make all amino acids besides leucine and tryptophan, and share these markers: $MAT\mathbf{a}\ can1$-$100\ hml\alpha\Delta$::$BLE\ leu9\Delta$::$KANMX6\ prACT1-yCerulean-tADH1@URA3$. Strain Leu$^{\text{FBR}}$ Trp$^-$ has the additional modifications $his3\Delta$::$prACT1-ymCitrine-tADH1$:$HIS3MX6\ LEU4^{\text{FBR}}\ trp2\Delta$::$NATMX4$, and strain Leu$^-$ Trp$^{\text{FBR}}$ has $his3\Delta$::$prACT1-ymCherry-tADH1$:$HIS3MX6\ leu4\Delta$::$HPHMX4\ TRP2^{\text{FBR}}$. The enzymes Leu4$^{\text{FBR}}$ (Ref. [31]) and Trp2$^{\text{FBR}}$ (Ref. [32]) are insensitive to feedback-inhibition by leucine and tryptophan, respectively. Both strains express the cyan fluorescent protein yCerulean. Leu$^{\text{FBR}}$ Trp$^-$ also expresses the yellow fluorescent protein mCitrine, and Leu$^-$ Trp$^{\text{FBR}}$ the red fluorescent protein mCherry. To enhance contrast, Leu$^{\text{FBR}}$ Trp$^-$ and Leu$^-$ Trp$^{\text{FBR}}$ are depicted as yellow and blue, respectively. More strain are in the SI Sec. S1.

**Growth conditions** We used 1% agarose plates with CSM-leucine-tryptophan (complete synthetic medium as described in Ref. [46], except 2 mg/l of adenine and no leucine and tryptophan were used), plus appropriate amounts of leucine and tryptophan. For fully complemented CSM, we added at least $1524\,\mu$M leucine and $196\,\mu$M tryptophan. Cells were pre-grown in liquid CSM at $30^\circ$C in exponential phase for more than 12 h, counted with a Beckman Coulter counter, and mixed in appropriate ratios. The mix was spun down and vortexed after discarding the supernatant. A $0.5\,\mu$l drop of the mix ($\approx 10^9$ cells/ml) was pipetted on agar plates that had dried for 2 days post-pouring. Plates were incubated at $30^\circ$C in a humidified box for 7 days and imaged with a Zeiss Lumar stereoscope.

**Radial growth velocity** Colonies were imaged once a day, and their radii, determined by circle fitting with MATLAB, were fitted with a straight line for days 4-7. The velocity is the average of slopes from at least 3 different colonies.

**Boundary fraction ($f$)** Cells were scraped from colony boundaries with a pipet tip, avoiding mutant sectors, and resuspended in PBS. The fraction $f$ of red fluorescent cells was determined on a Beckton-Dickinson LSR Fortessa flow cytometer.

**Patch width ($L$)** Using MATLAB, we determined the local maxima in the yellow fluorescence intensity (normalized by the cyan fluorescent intensity to correct effects of varying colony thickness and unequal lighting, and smoothed over 15 pixels) plotted along the circumference separately for each radius. The patch width $L$ is the circumference divided by twice the number of maxima. Data within 20 pixels from the colony boundary were excluded because of weak fluorescence intensity. Mutant sectors were excluded from the analysis. Using the red fluorescence intensity instead of the yellow intensity gave similar results.

**Histograms to characterize demixing** We constructed the histogram of the yellow fluorescence intensity (normalized by cyan) of 3000 individual pixels that were randomly selected from the region at 50 to $550\mu$m distance from the colony boundary, and determined the locations and number of its modes (separated by at least 3 bins). In a bootstrapping analysis, we performed this procedure 100 times to determine the average mode locations and the probability of observing only one mode. Using the red fluorescence intensity (normalized by cyan) gave similar results.

# 5 Acknowledgements

We thank Erik F. Hom, Kirill S. Korolev, and J. David van Dyken for discussions. Support for this work was provided by the National Institute of General Medical Sciences Grant P50GM068763 of the National Centers for Systems Biology, by the National Science Foundation through grant DMR-1005289, and by the Harvard Materials Research Science and Engineering Center through grant DMR-0820484. MJIM was supported by a research fellowship from the German Research Foundation and a grant from the National Philanthropic Trust.

# Genetic drift opposes mutualism during spatial population expansion - Supporting Information

Melanie J.I. Müller[1,2,3], Beverly I. Neugeboren[1,3], David R. Nelson[1,2,3], and Andrew W. Murray[1,3]

[1] FAS Center for Systems Biology, [2] Department of Physics, [3] Department of Molecular and Cellular Biology, Harvard University, Cambridge, MA, USA

# Contents

# S1 Experimental methods

This section gives additional details about our strains and experimental methods.

## S1.1 Strains

The strains used in this work are listed in Table S1. The mutualistic strains Leu$^{\text{FBR}}$ Trp$^-$ and Leu$^-$ Trp$^{\text{FBR}}$ are W303 strains that are prototrophic for all amino acids except for either leucine or tryptophan. They all have the genetic background W303 *MAT**a** can1-100 hmlα∆::BLE leu9∆::KANMX6 prACT1-yCerulean-tADH1@URA3*, and differ only at the *LEU4* and *TRP2* loci, as well as in their fluorescent marker at the *HIS3* locus (see below). The strains contain the S288C version of *BUD4* (standard W303 strains have a mutated *BUD4* [1], which causes alteration in the budding pattern and occasional failures in cell separation after cytokinesis). The deletion of the silent mating type locus *HMLα* prevents mating-type switching, and therefore mating, since all our strains are *MAT**a** at the mating type locus. The two strains thus behave essentially as two different species. Leu9 and Leu4 are isozymes for the first step in leucine biosynthesis [2], which makes the *leu9* deletion necessary in order to make a *leu4*-deletion auxotrophic for leucine. With the *mCerulean* gene under the actin promoter *prACT1*, all strains constitutively express a cyan fluorescent protein. We checked by flow cytometry that the cyan fluorescence intensity profiles were indistinguishable for all strains.

| Strain name | | Genotype at *LEU4* and *TRP2* loci | | fluorescent proteins |
|---|---|---|---|---|
| Leu$^{FBR}$ Trp$^-$ | yMM60 | *LEU4*$^{FBR}$ | *trp2Δ::NATMX4* | *ymCitrine, yCerulean* |
| Leu$^-$ Trp$^{FBR}$ | yMM65 | *leu4Δ::HPHMX4* | *TRP2*$^{FBR}$ | *ymCherry, yCerulean* |
| | yMM61 | *LEU4*$^{FBR}$ | *trp2Δ::NATMX4* | *ymCherry, yCerulean* |
| | yMM64 | *leu4Δ::HPHMX4* | *TRP2*$^{FBR}$ | *ymCitrine, yCerulean* |
| | yMM26 | *LEU4*$^{FBR}$ | *trp2Δ::NATMX4* | *ymCitrine* |
| | yMM31 | *leu4Δ::HPHMX4* | *TRP2*$^{FBR}$ | *ymCherry* |
| | yMM29 | *LEU4* | *trp2Δ::NATMX4* | *ymCherry* |
| | yMM32 | *leu4Δ::HPHMX4* | *TRP2* | *ymCitrine* |
| non-interacting | yJHK111 | *LEU4* | *TRP2* | *ymCitrine* |
| strains | yJHK112 | *LEU4* | *TRP2* | *ymCherry* |

Table S1: Strains used in this work. All strains have the genetic background *W303 MAT**a** can1-100*. All yMM strains have in addition *hmlαΔ::BLE leu9Δ::KANMX6*.
The fluorescent markers are incorporated as follows:
ymCitrine: *his3Δ::prACT1-ymCitrine-tADH1:HIS3MX6*,
ymCherry: *his3Δ::prACT1-ymCherry-tADH1:HIS3MX6*,
yCerulean: *prACT1-yCerulean-tADH1@URA3*.

The leucine overproducing strain Leu$^{FBR}$ Trp$^-$ is the strain yMM60, which has the (additional) genetic modifications *his3Δ::prACT1-ymCitrine-tADH1:HIS3MX6 LEU4$^{FBR}$ trp2Δ::NATMX4*. It constitutively expresses the yellow fluorescent protein *mCitrine* and is a tryptophan-auxotroph. The *LEU4$^{FBR}$* allele differs from the wild-type *LEU4* by the deletion of the codon 548; it is functional in leucine biosynthesis, but feedback-resistant (FBR) to inhibition by the end-product leucine. It was identified in Ref. [3] (named there *LEU4-1*). This strain is depicted as yellow (in accordance with its fluorescent protein) in the main text. The tryptophan overproducing strain Leu$^-$ Trp$^{FBR}$ is the strain yMM65, with *his3Δ::prACT1-ymCherry-tADH1:HIS3MX6 leu4Δ::HPHMX4 TRP2$^{FBR}$*. It constitutively expresses the red fluorescent protein *mCherry* and is a leucine-auxotroph. The $TRP2^{FBR}$ gene is the allele $TRP2 - S76L$, which was identified in Ref. [4] (designated $L_{76}$ in this reference) as a feedback-resistant, functional version of $TRP2$. This strain is depicted as blue in the main text (to enhance its contrast with the yellow color of the partner strain).

To rule out effects of the fluorescent proteins on the fitness or other phenotypes of the strains, we also used strains with 'swapped' fluorescent proteins: yMM61, with *his3Δ::prACT1-ymCherry-tADH1:HIS3MX6 LEU4$^{FBR}$ trp2Δ::NATMX4*, and yMM64, with *his3Δ::prACT1-ymCitrine-tADH1:HIS3MX6 leu4Δ::HPHMX4 TRP2$^{FBR}$*. In order to compare feedback-resistant and wild-type amino acid production we used the following strains: the FBR strains yMM26 and yMM31, which have the same genetic background as yMM60 and yMM65, respectively, except for being *URA3* wild-type at the Ura3 locus; and the WT strains yMM29 and yMM32, which are *LEU4 his3Δ::prACT1-ymCherry-tADH1:HIS3MX6* and *TRP2 his3Δ::prACT1-ymCitrine-tADH1:HIS3MX6*, respectively.

## S1.2 Fitness costs of amino acid overproduction due to feedback resistance

The relative fitness in direct competition in liquid culture was measured by a flow-cytometer-based competition assay as described in Ref. [5]. The relative fitness on plates was determined by analyzing the shape of sectors of the two competing strains in a colony as described in Ref. [6].

We first checked that the fluorescent proteins mCitrine and mCherry were neutral with respect to each other both in the liquid and the plate competition assays by verifying that the strains yMM60 and yMM61 as well as yMM63 and yMM64, which differ only in their fluorescent protein, have the same fitness in CSM (for medium definitions, see the Materials and Methods section of the main text).

We competed the mutualistic strains Leu$^{FBR}$ Trp$^-$ (yMM60) and Leu$^-$ Trp$^{FBR}$ (yMM65) in liquid CSM,

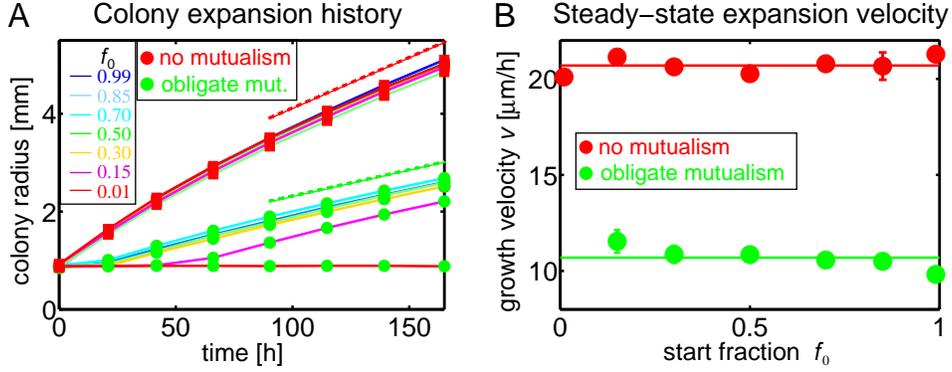

Figure S1: Colony expansion velocities. **A)** Radius increase over time for individual colonies of the mutualistic strains Leu$^-$ Trp$^{FBR}$ and Leu$^{FBR}$ Trp$^-$, inoculated at different start fractions $f_0$ of blue cells, under conditions of no (red dots) and obligate mutualism (green dots). In the no mutualism case, all colonies exhibit similar radius increases and reach a velocity of 20.7±0.4$\mu$m/h for times larger than 90h (upper red line). In contrast, obligately mutualistic colonies start expanding at different points in time: Colonies with start fraction $f_0 = 0.15$ start expanding at later times (solid purple line through green dots), and colonies with $f_0 = 0.01$ do not expand at all (solid red line through green dots). Expanding colonies reach a velocity of 10.7±0.6$\mu$m/h for times larger than 90h (upper green line), indicating that the mutualistic strains have reached steady-state expansion. **B)** Growth velocities for times larger than 90h are independent of the start fraction $f_0$ both for non-mutualistic (red circles) and for obligately mutualistic (green circles) colonies. Solid lines indicate the average velocities, which are the slopes of the corresponding straight upper red and green lines in (A).

and found that Leu$^-$ Trp$^{FBR}$ had a 2% fitness advantage compared to Leu$^{FBR}$ Trp$^-$. The same was true in a competition of the strains yMM61 and yMM64 with 'swapped' fluorescent proteins. This fitness difference is entirely due to leucine feedback-resistance, as the Trp$^{FBR}$ strain yMM31 and the Trp$^{WT}$ strain yMM32 have the same fitness, while the Leu$^{FBR}$ strain yMM26 and the Leu$^{WT}$ strain yMM29 also have a 2% fitness difference. Presumably, leucine overproduction due to feedback-resistance is more costly than tryptophan overproduction because leucine is more abundant than tryptophan, with ∼10% of yeast amino acids being leucine and only ∼1% being tryptophan [7–9].

### S1.3 Radial expansion velocities

**Radial expansion velocities.** When a yeast colony expands on an agar surface, its radius increases linearly with time after an initial transient of about 4 days, see Fig. S1A. In the linear regime, we characterize colony growth by an expansion velocity, defined as the slope of the radius versus time. For both mutualistic and non-mutualistic colonies, the steady-state expansion velocity is independent of its initial conditions, such as the inoculation volume and density or the initial mixing ratio of the blue and yellow strains.

**Initial transient in radius increase.** In contrast, the transient towards the linear regime can depend on the initial condition, since it takes longer to reach steady-state growth mode when the initial conditions are further away from it. As shown in Figure 3 in the main text, obligately mutualistic colonies are driven towards an optimal fraction $f^* = 0.5$ of blue cells. Thus, when colonies are inoculated with a start fraction $f_0$ close to $f^*$, they approach the steady-state growth velocity faster than when inoculated with a start fraction further from $f^*$. This effect is particularly noticeable for small $f_0$, see the orange and red data in Fig. S1A, i.e. when there is a very small fraction of leucine requiring cells. In the case of $f_0 = 0.01$, the colony never grew on experimental timescales.

**Growth as function of the amino acid concentrations.** Figure 2A in the main text shows that, as a function of a required amino acid, the growth velocity increases linearly before it saturates at a plateau. This data is reproduced in Fig. S2 (blue diamonds in (A), yellow squares in (B)). In addition, this figure shows that the growth velocities do not depend on the concentration of the non-required amino acid (yellow squares in (A), blue diamonds in (B)). The maximal growth velocities (plateau values) for Leu$^-$ Trp$^{FBR}$ and

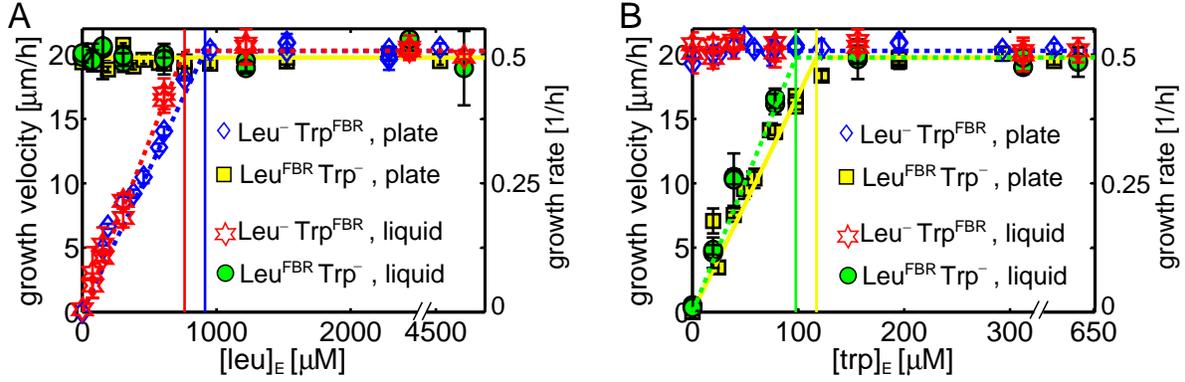

Figure S2: Comparison of growth velocities on plates and growth rates in liquid of Leu$^{\text{FBR}}$ Trp$^-$ (yellow squares for velocities and green circles for rates) and Leu$^-$ Trp$^{\text{FBR}}$ (blue diamonds for plates and red stars for liquid) for (A) varying leucine and (B) varying tryptophan concentrations. Growth velocities and rates are independent of non-required amino acids, but increase linearly with a required amino acid concentration before they saturate at a plateau. When the liquid growth rates (red stars for Leu$^-$ Trp$^{\text{FBR}}$ and green circles for Leu$^{\text{FBR}}$ Trp$^-$) are scaled by a factor of $b_{vg} = 40\,\mu\text{m}$, they are very similar to the growth velocities on plates (blue diamonds for Leu$^-$ Trp$^{\text{FBR}}$ and yellow squares for Leu$^{\text{FBR}}$ Trp$^-$). However, crossovers from linear to plateau regimes for liquid growth occur at concentrations [leu]$_c$ = 762 $\mu$M (red vertical line) and [trp]$_c$ = 97.6 $\mu$M (green vertical line) that are lower by a factor of $1 + \kappa = 1.18$ than the corresponding crossover concentrations (blue and yellow vertical lines) for the plate growth velocities, as explained in Section S2.2.

Leu$^{\text{FBR}}$ Trp$^-$ are $v_{\text{L}} = 20.3\,\mu\text{m/h}$ and $v_{\text{T}} = 19.8\,\mu\text{m/h}$, respectively. The 2.5% difference in the plateau growth rates agrees, within experimental error, with the difference in fitness between the two strains as measured by the competitive fitness assay described in Section S1.2.

We also measured the growth rate in well-mixed culture as function of the external amino acid concentration. As shown in Fig. S2, the growth rates behave very similarly to the growth velocities. In fact, when scaled by the same factor of $b_{vg} = 40\,\mu\text{m}$, the growth rate plateaus for both Leu$^-$ Trp$^{\text{FBR}}$ and Leu$^{\text{FBR}}$ Trp$^-$ overlap with the growth velocity plateaus. The growth rates in the linear regimes are similar but not identical, since the growth rates in liquid increase slightly faster than the growth velocities on solid medium. This leads to higher crossover concentrations from the linear to the plateau regime for plate growth (blue and yellow vertical lines) than for liquid growth (red and green vertical lines). We normalize the leucine and tryptophan concentrations such that the crossover concentrations are at $n_{\text{EL}} = [\text{leu}]/[\text{leu}]_c = 1$ and $n_{\text{ET}} = [\text{trp}]/[\text{trp}]_c = 1$ for liquid growth, with [leu]$_c$ = 762 $\mu$M and [trp]$_c$ = 97.6 $\mu$M, since the liquid concentrations reflect the concentrations felt by the cells. The slower growth on plates is due to an amino acid depletion layer around the colony and leads to crossover concentrations which are larger by a factor of 1.18, see also Section S2.2.

### S1.4 Determination of yeast colony growth parameters

This subsection summarizes how we determined the model parameters listed in Table S2.

**Cellular diffusion constant $D_s$ and genetic diffusion constant $D_g$.** We determined the spatial diffusion constant $D_s$ (for cellular diffusion due to the jostling of cell division) and the genetic diffusion constant $D_g$ (which characterizes the strength of genetic drift) with the sectoring assay described in Ref. [10]. When performing this assay with the non-interacting equally fit strains yJHK111 and yJHK112, we obtained $D_s/\tau_g = 15\,\mu\text{m}^2/\text{h}$ and $D_g/\tau_g = 1.3\,\mu\text{m/h}$, where $\tau_g = 1.5\,\text{h}$ is the yeast generation time. The value of $D_s$ for cellular diffusion lies in the expected range: In each generation, cells push by a diameter of $a = 5\,\mu\text{m}$, so we expect the diffusion constant to be of the order of $D_s = a^2 = 25\,\mu\text{m}^2$ per generation, or $D_s/\tau_g = 17\,\mu\text{m}^2/\text{h}$. The value of the genetic diffusion constant $D_g$ allows us to estimate the effective population density $\rho_e \approx 1/D_g$ [10]. The resulting effective density $\rho_e \approx 0.5/\mu\text{m}$ of about 2-3 cells in a region of the size of a cell is rather low, especially when considering that yeast colonies pile up to a height of ∼ 100 cells. This result is similar to even lower effective population densities for the bacteria *E. coli* and *P. aeruginosa* observed in Ref. [10], and implies that the dynamics of the boundaries between different genotypes are dominated by the behavior of the cells at the extreme edge of the growing colonies.

**Size $b$ of the active layer.** To estimate the size $b$ of the active layer, within which cell proliferation drives radial growth, we use the observation that the maximal colony growth velocity $v$ is $b_{vg} = 40\,\mu$m times the maximal growth rate $g$ in liquid culture (Section S1.3). Within a generation time $\tau_g \sim 1/g$, the number of cells increases by $\Delta N = Ubc$, where $U$ is the colony circumference and $c$ is the cellular surface carrying capacity. In the same time $\tau_g$, the colony increases in area by $\Delta A = Uv\tau_g = Uv/g$, which corresponds to $\Delta N = c\Delta A$ new cells. Thus, $v/g = b$. With $v/g = b_{vg} = 40\,\mu$m, we obtain the reasonable estimate of $b = 40\,\mu$m for the size of the active layer [11, 12].

**Surface carrying capacity $c$.** Colonies grown between 4 and 7 days (to obtain a variety of sizes, and to measure in the same time window used for all other experiments) were imaged with a Zeiss Lumar Stereoscope to determine the colony area by image analysis (circle fit with MATLAB). Directly after imaging, a small agar pad with the colony on it was cut out of the plate and suspended in a known amount of PBS. All cells were washed off the agar by strong vortexing. The cell concentration in the PBS was then determined by using a Beckman Coulter counter, from which the total number of cells in the colony could be calculated. The surface carrying capacity was determined as the slope of the cell number versus colony area, with the result of $10\,\text{cells}/\mu\text{m}^2$. Taking into account that yeast colonies have a height of $0.5 - 1$ mm, this surface density corresponds to a volume density of about $0.01\,\text{cells}/\mu\text{m}^3$, which is similar to dense spherical packing of yeast cells of diameter $5\,\mu$m. Since we are interested in the cell density at the colony boundary, and since yeast colonies are higher in the middle than at the boundary, we take $c = 5\,\text{cells}/\mu\text{m}^2$ [11].

**Amino acid diffusion constant $D_a$.** Diffusion through the agar medium is similar to diffusion in water, since the agar pore size of about 500Å [13] is much larger than the amino acid hydrodynamic radius of about 3Å [14]. The diffusion constant for the similar-sized leucine and tryptophan molecules is $D_a = 3\,\text{mm}^2/\text{h}$ [15].

**Diffusive flux rate $d$.** The derivation of $d = 5/\text{h}$ is described in Section S5.

**Amino acid crossover concentrations $[\text{leu}]_c$ and $[\text{trp}]_c$.** The leucine and tryptophan crossover concentrations between the regime where growth depends linearly on the amino acid concentrations and the regime where growth is saturated are determined in Fig. S2 as $[\text{leu}]_c = 762\,\mu\text{M} \,\hat{=}\, 2.3 \cdot 10^9\,\text{molecules}/\mu\text{m}^2$ and $[\text{trp}]_c = 97.6\,\mu\text{M} \,\hat{=}\, 2.9 \cdot 10^8\,\text{molecules}/\mu\text{m}^2$. Here, we have used the agar height of 5 mm to transform the three-dimensional volume concentrations to two-dimensional surface concentrations, i.e. to the number of molecules available to cells from the agar directly beneath.

**Amino acid uptake rates.** From the literature, we estimate the amino acid uptake rates as $k_L = 10^5\,\text{molecules/s/cell}$ for leucine [8, 16, 17] and $k_T = 1.5 \cdot 10^4\,\text{molecules/s/cell}$ for tryptophan [8, 18]. These values are in agreement with the value of the parameter $\kappa = k_L\,c/[\text{leu}]_c/d = k_T\,c/[\text{trp}]_c/d = 0.18$ (determined independently from our growth rate and velocity measurements in Fig. S2 in Section S2.2) and our above estimates of $[\text{leu}]_c$, $[\text{trp}]_c$, $c$ and $d$.

**Amino acid secretion rates.** We expect the amino acid leakage rates to be proportional to the internal cellular amino acid concentrations. Since yeast cells contain an order of magnitude more leucine than tryptophan [7–9], we expect the leucine secretion rate to be an order of magnitude larger that the tryptophan secretion rate. From our experimental result that our yeast strains grow twice as fast under non-mutualistic than under obligate mutualistic conditions (Section S3.1), we determined the parameter $\rho = 1.09$ from the equation $\rho = 1 + \kappa/2$. With $\rho = r_L\,c/[\text{leu}]_c/d = r_T\,c/[\text{trp}]_c/d$, we obtain the leucine and tryptophan secretion rates $r_L = 7 \cdot 10^5\,\text{molecules/s/cell}$ and $r_T = 9 \cdot 10^4\,\text{molecules/s/cell}$. These values are of the order of magnitude of amino acid secretion rates measured in Ref. [19], but are on the larger side, presumably because our strains are overproducing leucine and tryptophan due to the engineered feedback-resistance and thus leak large amounts of excess amino acids.

## S2 Modeling growth of the mutualistic yeast strains in the presence of amino acids

In this section we describe a model for the growth of our mutualistic yeast strains in the presence of amino acids in more detail than in the main text. The model parameters are listed in Table S2.

| Symbol | Definition | Value | Description |
|---|---|---|---|
| $\kappa$ | $\frac{k_\mathrm{L}}{[\mathrm{leu}]_\mathrm{c}}\frac{c}{d} = \frac{k_\mathrm{T}}{[\mathrm{trp}]_\mathrm{c}}\frac{c}{d}$ | 0.18 | reduced amino acid uptake rate |
| $\rho$ | $\frac{r_\mathrm{L}}{[\mathrm{leu}]_\mathrm{c}}\frac{c}{d} = \frac{r_\mathrm{T}}{[\mathrm{trp}]_\mathrm{c}}\frac{c}{d}$ | 1.09 | reduced amino acid secretion rate |
| $l_\mathrm{a}$ | $\sqrt{\frac{D_\mathrm{a}/d}{1+\kappa}}$ | $710\,\mu\mathrm{m}$ | length scale of amino acid depletion |
| $l_\mathrm{b}$ | $\frac{2D_\mathrm{s}}{b}$ | $0.83\,\mu\mathrm{m}$ | length scale of patch boundary diffusion |
| $s_\mathrm{c}$ | $D_\mathrm{g}^2/D_\mathrm{s}$ | 0.18 | critical selection strength to overcome genetic drift |

Table S2: Parameters used in the theoretical calculations of Section S3 and S4. As given by the definitions in the table, these 'reduced' parameters are compounds of directly biologically relevant parameters: the leucine and tryptophan uptake rates $k_\mathrm{L}$ and $k_\mathrm{T}$, the leucine secretion rates $r_\mathrm{L}$ and $r_\mathrm{T}$, the leucine and tryptophan crossover concentrations $[\mathrm{leu}]_\mathrm{c}$ and $[\mathrm{trp}]_\mathrm{c}$, the cellular surface carrying capacity $c$, the amino acid diffusive flux rate $d$, the amino acid diffusion constant $D_\mathrm{a}$, the size $b$ of the actively growing layer, the cellular diffusion constant $D_\mathrm{s}$ and the genetic diffusion constant $D_\mathrm{g}$. The determination of the parameter values is described in Section S1.4.

## S2.1 Model for amino acid dynamics

**Growth velocities.** As shown in Section S1.3, the growth rates and velocities increase roughly linearly with the concentration of the required amino acid until they saturate at a plateau. This motivates us to write the growth velocities as function of the rescaled amino acid concentrations $n_\mathrm{L} = [\mathrm{leu}]/[\mathrm{leu}]_\mathrm{c}$ and $n_\mathrm{T} = [\mathrm{trp}]/[\mathrm{trp}]_\mathrm{c}$:

$$\mathrm{Leu^- Trp^{FBR}\ expansion\ velocity:\ } V_\mathrm{L}(n_\mathrm{L}) = v_\mathrm{L}\, T(n_\mathrm{L}), \tag{1}$$
$$\mathrm{Leu^{FBR} Trp^-\ expansion\ velocity:\ } V_\mathrm{T}(n_\mathrm{T}) = v_\mathrm{T}\, T(n_\mathrm{T}), \tag{2}$$

where we have defined the threshold function:

$$T(x) = \begin{cases} x & \text{for } x \leq 1 \\ 1 & \text{for } x > 1 \end{cases} \tag{3}$$

Note that the velocities are expressed as function of concentrations $n_\mathrm{L}$ and $n_\mathrm{T}$ felt by the cells, which can differ from the external amino acid concentrations $n_\mathrm{EL}$ and $n_\mathrm{ET}$ with which the agar medium was prepared, see below.

**Amino acid uptake.** Under steady state growth conditions, the uptake of amino acids that are not produced by the cell must equal the amount used in the production of cellular biomass. Thus, the amino acid uptake rates must be proportional to the growth rates:

$$\mathrm{Leu\ uptake\ rate\ of\ Leu^- Trp^{FBR}:}\quad K_\mathrm{L}(n_\mathrm{L}) = k_\mathrm{L}\, T(n_\mathrm{L}) \tag{4}$$
$$\mathrm{Trp\ uptake\ rate\ of\ Leu^{FBR} Trp^-:}\quad K_\mathrm{T}(n_\mathrm{T}) = k_\mathrm{T}\, T(n_\mathrm{T}) \tag{5}$$

where $k_\mathrm{L}$ and $k_\mathrm{T}$ are the maximal leucine and tryptophan uptake rates, respectively.

**Amino acid secretion.** Sine our yeast strains are feedback-resistant in the production of leucine or tryptophan, they cannot regulate production in response to the available amount of the end product. We therefore assume that yeast secrete amino acids at a constant rate,

$$\mathrm{Leucine\ secretion\ rate\ of\ Leu^{FBR} Trp^-:}\quad R_\mathrm{L}(n_\mathrm{L}) = r_\mathrm{L}, \tag{6}$$
$$\mathrm{Tryptophan\ secretion\ rate\ of\ Leu^- Trp^{FBR}:}\quad R_\mathrm{T}(n_\mathrm{T}) = r_\mathrm{T}, \tag{7}$$

see also Ref. [20].

**Amino acid dynamics.** The dynamics of leucine and tryptophan with concentrations $[\mathrm{leu}]$ and $[\mathrm{trp}]$

can be described by

$$\frac{\partial}{\partial t}[\text{leu}] = r_\text{L}\, c_\text{T} - K_\text{L}([\text{leu}])\, c_\text{L} + D_\text{a}\, \Delta[\text{leu}], \quad (8)$$

$$\frac{\partial}{\partial t}[\text{trp}] = r_\text{T}\, c_\text{L} - K_\text{T}([\text{trp}])\, c_\text{T} + D_\text{a}\, \Delta[\text{trp}]. \quad (9)$$

The first and second term describe amino acid secretion and uptake, respectively, by Leu$^-$ Trp$^{\text{FBR}}$ cells (surface concentration $c_\text{L}$) and Leu$^{\text{FBR}}$ Trp$^-$ cells (surface concentration $c_\text{T}$). The last term describes diffusion of amino acids with diffusion constant $D_\text{a}$.

**Yeast colony growth resembles a chemostat.** As described in more detail in Section S5, only cells in a small active layer near the colony boundary are actively growing. The growth dynamics within the active layer in the reference frame of the boundary is reminiscent of growth in a chemostat, in which nutrient are added at a constant rate that is balanced by outflux of waste (medium and cells) [21]. For the active layer of a colony, there is continuous influx of nutrients (e.g. glucose) due to diffusive flux towards the colony. Cells are removed 'behind' the active layer, because cells in the colony interior are not growing and therefore effectively 'gone'. When the colony grows at constant velocity, nutrient influx and cell 'outflux' are balanced, so that the number of actively growing cells remains constant. Thus, we can describe cell dynamics in terms of the fraction

$$f = \frac{c_\text{L}}{c_\text{L} + c_\text{T}} \quad (10)$$

of Leu$^-$ Trp$^{\text{FBR}}$ cells of the constant surface carrying capacity $c = c_\text{L} + c_\text{T}$. As explained in Section S5, chemostat-like growth also means that the Laplace operator $\Delta$ can be decomposed (in a frame co-moving with the frontier) into angular and radial diffusion in a specific way, e.g. for leucine

$$\Delta[\text{leu}] = \frac{\partial^2}{\partial x^2}[\text{leu}] - d\left([\text{leu}] - [\text{leu}]_\text{E}\right). \quad (11)$$

Here, the first term describes angular diffusion along the front coordinate $x$, with $\partial x = R\partial\phi$. The second term replaces radial diffusion by effectively describing the radial diffusive flux into or out of the colony as dilution with rate $d$ into medium with concentration $[\text{leu}]_\text{E}$ (the concentration in the agar medium far away from the colony, i.e. the concentrations with which the medium was prepared). With Equations (10), (11), the diffusion equations (8,9) become:

$$\frac{\partial[\text{leu}]}{\partial t} = D_\text{a}\frac{\partial^2[\text{leu}]}{\partial x^2} - d\left([\text{leu}] - [\text{leu}]_\text{E}\right) + r_\text{L}\, c\, (1-f) - K_\text{L}([\text{leu}])\, c\, f \quad (12)$$

$$\frac{\partial[\text{trp}]}{\partial t} = D_\text{a}\frac{\partial^2[\text{trp}]}{\partial x^2} - d\left([\text{trp}] - [\text{trp}]_\text{E}\right) + r_\text{T}\, c\, f - K_\text{T}([\text{trp}])\, c\, (1-f) \quad (13)$$

**Non-dimensionalization.** These equations can be written in a non-dimensionalized form for the rescaled amino acid concentrations $n_\text{L} = [\text{leu}]/[\text{leu}]_\text{c}$ and $n_\text{T} = [\text{trp}]/[\text{trp}]_\text{c}$ by expressing time in units of the inverse diffusive rate $d$ and space in units of the diffusion length scale $\sqrt{D_\text{a}/d}$:

$$\frac{\partial n_\text{L}}{\partial \tilde{t}} = \frac{\partial^2 n_\text{L}}{\partial \tilde{x}^2} - (n_\text{L} - n_\text{EL}) + \rho_\text{L}(1-f) - \kappa_\text{L}\, T(n_\text{L})f \quad (14)$$

$$\frac{\partial n_\text{T}}{\partial \tilde{t}} = \frac{\partial^2 n_\text{T}}{\partial \tilde{x}^2} - (n_\text{T} - n_\text{ET}) + \rho_\text{T} f - \kappa_\text{T}\, T(n_\text{T})(1-f) \quad (15)$$

with $\tilde{t} = t\, d$ and $\tilde{x} = x/\sqrt{D_\text{a}/d}$. The secretion and uptake parameters are now dimensionless:

$$\rho_\text{L} \equiv \frac{r_\text{L}\, c}{[\text{leu}]_\text{c}\, d}, \qquad \rho_\text{T} \equiv \frac{r_\text{T}\, c}{[\text{trp}]_\text{c}\, d}, \qquad \kappa_\text{L} \equiv \frac{k_\text{L}\, c}{[\text{leu}]_\text{c}\, d}, \qquad \kappa_\text{T} \equiv \frac{k_\text{T}\, c}{[\text{trp}]_\text{c}\, d}. \quad (16)$$

Intuitively, in the reduced parameter $r_\text{L}\, c/([\text{leu}]_\text{c}\, d)$, the 'free' secretion rate $r_\text{L}$ is reduced by the factor $d$ due to diffusive loss of amino acids from the colony. In addition, it is rescaled from cellular to amino acid concentrations with the 'conversion factor' $c/[\text{leu}]_\text{c}$.

Note that for our mutualistic yeast strains, the tryptophan secretion rate $r_T$, uptake rate $k_T$, and crossover concentration $[\text{trp}]_c$ are all lower than the corresponding quantities for leucine by a factor of 8, see Section S1.4. Since all dimensionless rates depend on the ratio of a secretion or uptake rate and the crossover concentration, this means that the dimensionless secretion and uptake parameters have the same values for the two strains,

$$\text{rescaled nutrient uptake rate:} \quad \kappa \equiv \kappa_L = \kappa_T, \tag{17}$$

$$\text{rescaled nutrient secretion rate:} \quad \rho \equiv \rho_L = \rho_T. \tag{18}$$

The numerical values of these dimensionless parameters are listed in Table S2.

## S2.2 Steady-state amino acid concentrations

**Steady-state amino acid concentrations.** The cell concentrations $c_L$ and $c_T$ vary in angular direction due to spatial demixing into patches or sectors. If only one cell type is present, or on length scales large compared to the patch or sector width, the cells in the active layer are essentially well-mixed, and the problem becomes radially symmetric, i.e. independent of the radial angle $\phi$. In this case the amino acid dynamics of Equations (14) and (15) become

$$\frac{1}{d}\frac{\partial}{\partial t} n_L = \rho(1-f) - \kappa T(n_L) f - (n_L - n_{EL}), \tag{19}$$

$$\frac{1}{d}\frac{\partial n_T}{\partial t} = \rho f - \kappa T(n_T)(1-f) - (n_T - n_{ET}). \tag{20}$$

For a calculation that takes into account cellular patches along the colony front, see Section S4.

Since the nutrient dynamics are fast compared to colony growth, we can assume that the cell fraction $f$ is constant on nutrient time scales. The nutrient dynamics equations Equations (14) and (15) then have the quasi-stationary solution

$$n_L(f) = \begin{cases} \frac{n_{EL} + \rho(1-f)}{1+\kappa f} \equiv n_{mL}(f) & \text{for } f \geq f_{C1} \quad (\text{case } n_L \leq 1) \\ n_{EL} + \rho(1-f) - \kappa f \equiv n_{dL}(f) & \text{for } f \leq f_{C1} \quad (\text{case } n_L \geq 1) \end{cases} \tag{21}$$

$$n_T(f) = \begin{cases} \frac{n_{ET} + \rho f}{1+\kappa(1-f)} \equiv n_{mT}(f) & \text{for } f \leq f_{C2} \quad (\text{case } n_T \leq 1) \\ n_{ET} + \rho f - \kappa(1-f) \equiv n_{dT}(f) & \text{for } f \geq f_{C2} \quad (\text{case } n_T \geq 1) \end{cases} \tag{22}$$

with the limiting allele frequencies

$$\begin{aligned} f_{C1} &= \frac{\rho - 1 + n_{EL}}{\rho + \kappa} = \frac{1}{2} + \frac{2n_{EL} - (2+\kappa-\rho)}{2(\rho+\kappa)}, \\ f_{C2} &= \frac{\kappa + 1 - n_{ET}}{\rho + \kappa} = \frac{1}{2} - \frac{2n_{ET} - (2+\kappa-\rho)}{2(\rho+\kappa)}. \end{aligned} \tag{23}$$

In view of Equations (1) and (2), this means that the growth velocities of the two strains are given by:

$$V_L(f) = \begin{cases} v_L\, n_{mL}(f) & \text{for } f \geq f_{C1} \\ v_L & \text{for } f \leq f_{C1} \end{cases}, \quad \text{and} \quad V_T(f) = \begin{cases} v_T\, n_{mT}(f) & \text{for } f \leq f_{C2} \\ v_T & \text{for } f \geq f_{C2} \end{cases} \tag{24}$$

**Crossover concentrations from linear to saturated growth.** For the special case of single-strain colonies, consisting e.g. of only Leu$^-$ Trp$^{\text{FBR}}$ cells ($f = 1$), the steady-state leucine concentration is

$$n_L = \begin{cases} \frac{n_{EL}}{1+\kappa} & \text{for } n_{EL} \leq 1 + \kappa, \\ n_{EL} - \kappa & \text{for } n_{EL} \geq 1 + \kappa, \end{cases} \tag{25}$$

This means that the colony expansion velocity from Equation (24) as function of the external amino acid concentration is

$$V_{\text{colony}} = V_L(n_{EL}) = \begin{cases} v_L \frac{n_{EL}}{1+\kappa} & \text{for } n_{EL} \leq 1 + \kappa, \\ v_L & \text{for } n_{EL} \geq 1 + \kappa, \end{cases} \tag{26}$$

Thus, when measuring the colony velocity as function of the external amino acid concentration, as in Section S1.3, the velocity passes from the linear regime (proportional to the amino acid concentration) to the saturated regime (plateau velocity) at $n_{\mathrm{EL}} = 1 + \kappa$. This threshold is higher than the crossover value of 1 naively expected from Equation (1) because the amino acid concentration $n_{\mathrm{L}}$ felt by the colony is lower than the external concentration by the factor $(1 + \kappa)$. Similar equations hold for single-strain colonies of $\mathrm{Leu}^{\mathrm{FBR}} \mathrm{Trp}^-$ cells. In particular, the factor $(1 + \kappa)$ is the same for both strains, which we indeed observe experimentally (Fig. S2).

## S3 Generalized Moran model for mutualism at the front

**Allele frequency dynamics.** The expansion of a circular yeast colony on an agar surface is a two-dimensional process. However, since only cells in a small active layer near the colony boundary are actively growing (Section S5) one can approximate the expansion by a one-dimensional process along the front coordinate $x$ [22]. In addition, because colony front growth is similar to growth in a chemostat (Section S2.1), the effective population size remains approximately constant as long as the inflationary effect of the radius increase is not too large. In the context of population genetics, our two different yeast strains can be viewed as representing two different species or as two different alleles. In total, the selection dynamics for our two 'alleles' can be described within the framework of the Moran model, with the addition of a frequency-dependent growth rate and spatial diffusion [22–24]:

$$\frac{\partial f}{\partial \tau} = D_{\mathrm{s}} \frac{\partial^2 f}{\partial x^2} + s(f) f(1-f) + \sqrt{\widetilde{D_{\mathrm{g}}}(f) f(1-f)} \, \Gamma(x, \tau), \tag{27}$$

where $f(x, \tau)$ is the fraction of $\mathrm{Leu}^- \mathrm{Trp}^{\mathrm{FBR}}$ cells as function of the coordinate $x$ along the front and the time $\tau$ measured in generations. The first term on the r.h.s. of Equation (27) describes spatial diffusion with diffusion constant $D_{\mathrm{s}}$, which arises because dividing yeast cells push neighboring cells around. The second term describes selection with the selection coefficient

$$s(f) \equiv \frac{V_{\mathrm{L}}(f) - V_{\mathrm{T}}(f)}{f V_{\mathrm{L}}(f) + (1-f) V_{\mathrm{T}}(f)}. \tag{28}$$

Selection arises due to differences in the growth velocities $V_{\mathrm{L}}$ and $V_{\mathrm{T}}$ of the two strains, as given by Equation (24). Since the growth velocities depend on the 'allele frequency' $f$ because of the mutualistic interaction, the selection coefficient $s$ is frequency-dependent. The last term in Equation (27) describes genetic drift, i.e. noise due to cell number fluctuations. $\Gamma = \Gamma(x, \tau)$ is an Itō delta-correlated Gaussian noise, and the genetic diffusion constant

$$\widetilde{D_{\mathrm{g}}}(f) = D_{\mathrm{g}} \frac{V_{\mathrm{L}}(f) + V_{\mathrm{T}}(f)}{2 \left( f V_{\mathrm{L}}(f) + (1-f) V_{\mathrm{T}}(f) \right)} \tag{29}$$

characterizes the strength of the genetic drift. It is frequency-dependent due to its dependency on the growth velocities. However, its frequency-dependence is weak compared to the frequency-dependence of $s(f)$ so that we use $\widetilde{D_{\mathrm{g}}}(f) \approx D_{\mathrm{g}}$.

### S3.1 The effect of mutualistic selection

We first examine the effect of the mutualistic interaction which is incorporated in the selection coefficient $s(f)$. If the selection term dominates in Equation (27), for example in a well-mixed culture with high cell numbers (so $D_{\mathrm{g}} \approx 0$), the dynamics become

$$\frac{\partial f}{\partial \tau} = s(f) f(1-f). \tag{30}$$

As shown above, amino acid secretion and uptake is symmetric for our two strains because of $\kappa = \kappa_{\mathrm{L}} = \kappa_{\mathrm{T}}$ and $\rho = \rho_{\mathrm{L}} = \rho_{\mathrm{T}}$ (Equations (17) and (18)). For simplicity, we neglect here the 2% difference in growth velocities and use $v = v_{\mathrm{L}} = v_{\mathrm{T}}$. (We will consider the effect of the small growth velocity difference later in Section S3.2.) With these assumptions, the selection coefficient can be obtained from the steady-state amino acid concentrations (21,22) via the growth velocities (24). Its form depends on which of the limiting allele frequencies $f_{C1}$ and $f_{C2}$ given in Equation (23) is larger,

$$f_{C1} \leq f_{C2} \quad \Leftrightarrow \quad n_{EL} + n_{ET} \leq 2 + \kappa - \rho. \tag{31}$$

In the case $f_{C1} \leq f_{C2}$, the selection coefficient is:

$$s(f) = \begin{cases} \frac{1-n_{mT}(f)}{f+(1-f)n_{mT}(f)} & \text{for} \quad 0 \leq f \leq f_{C1} \quad (\text{regime } C_T \text{ with } n_L > 1, n_T < 1) \\ \frac{n_{mL}(f)-n_{mT}(f)}{fn_{mL}(f)+(1-f)n_{mT}(f)} & \text{for} \quad f_{C1} \leq f \leq f_{C2} \quad (\text{regime M with } n_L < 1, n_T < 1) \\ \frac{n_{mL}(f)-1}{fn_{mL}(f)+(1-f)n_{mT}} & \text{for} \quad f_{C2} \leq f \leq 1 \quad (\text{regime } C_L \text{ with } n_L < 1, n_T > 1) \end{cases} \tag{32}$$

and in the opposite case $f_{C1} \leq f_{C2}$:

$$s(f) = \begin{cases} \frac{1-n_{mT}(f)}{f+(1-f)n_{mT}(f)} & \text{for} \quad 0 \leq f \leq f_{C2} \quad (\text{regime } C_T \text{ with } n_L > 1, n_T < 1) \\ 0 & \text{for} \quad f_{C2} \leq f \leq f_{C1} \quad (\text{regime N with } n_L > 1, n_T > 1) \\ \frac{n_{mL}(f)-1}{fn_{mL}(f)+(1-f)n_{mT}} & \text{for} \quad f_{C1} \leq f \leq 1 \quad (\text{regime } C_L \text{ with } n_L < 1, n_T > 1) \end{cases} \tag{33}$$

In these equations, we have labeled the frequency region in which both strains are amino-acid limited ($n_L, n_T < 1$) as M for 'mutualistic', since in this regime both strains benefit from the additional amino acids due to the mutualistic interaction. This contrasts with the 'neutral' N region in which both strains are saturated for their required amino acids, $n_L, n_T > 1$, so that there is no benefit due to mutualism. The 'commensal' regions $C_T$ and $C_L$ denote the cases ($n_L > 1, n_T < 1$) and ($n_L < 1, n_T > 1$), respectively, in which one strain is amino-acid limited but the other is not. Some of these regions may vanish depending on whether the limiting fractions $f_{C1}$ and $f_{C2}$ are in the relevant interval $(0, 1)$:

$$0 \leq f_{C1} \leq 1 \quad \Leftrightarrow \quad 1 - \rho \leq n_{EL} \leq 1 + \kappa, \tag{34}$$

$$1 \geq f_{C2} \geq 0 \quad \Leftrightarrow \quad 1 - \rho \leq n_{ET} \leq 1 + \kappa. \tag{35}$$

**'Phase diagram'.** In total, the selection coefficient $s(f)$ depends on the parameters $\kappa$, $\rho$, and the external amino acid concentrations $n_{EL}, n_{ET}$. Depending on the values of the allele frequencies $f_{C1}$ and $f_{C2}$ (that mark the transitions between the commensal and the mutualistic or neutral regimes of the selection coefficient) in Equations (31,34,35), $s(f)$ could be in one of ten possible 'phases', M, $C_TM$, $MC_L$, $C_TMC_L$, $C_TNC_L$, $C_TN$, $NC_L$, $C_T$, $C_L$, and N. Here, for example, the notation $C_TM$ means that $s(f)$ has the functional form of the $C_T$ regime (tryptophan limiting, leucine not) in Equation (31) for $0 < f < f_{C1}$ and the functional form of the M regime (tryptophan and leucine both limiting) in Equation (31) for $f_{C1} < f < 1$. As we will see below, for yeast parameters, only five of the ten possible phases appear; they are shown in the $(n_{EL}, n_{ET})$ plane in Fig. S3A. To interpret the graphs of the selection term $s(f) f(1-f)$ as function of the allele frequency $f$, note that positive values of the selection term mean that selection acts to increase $f$, while negative values drive towards smaller $f$. A selection term of zero, $s(f) f(1-f) = 0$, defines a fixed point of the selection dynamics. The fixed point is stable if the selection term is positive to the left and negative to the right of it. Thus, in the $C_TMC_L$ phase, selection drives the system towards a stable fixed point within the M regime (unless $f = 0$ or $f = 1$, which are absorbing boundaries of the system). In the $C_TNC_L$, $C_TN$ and $NC_L$ phases, selection drives the system from the commensal regimes $C_T$ or $C_L$ towards the neutral N regime where the selective force is zero. If genetic drift is added, it will dominate in this neutral regime, causing random fluctuations of the allele frequency $f$.

The phase diagram in Fig. S3A shows that for all amino acid concentrations below the crossover concentration, $n_{EL}, n_{ET} < 1 + \kappa$, selection acts towards the interior of the frequency interval $(0, 1)$ and thus promotes mixing (red area in Fig. S3(A)). For very small amino acid concentrations (dark red triangle), selection drives the system towards a well-defined stable frequency $f^*$, while for intermediate concentrations (light red area), selection only drives towards a neutral region within the interval $(0, 1)$. For larger concentrations, $n_{EL}, n_{ET} > 1 + \kappa$, there is no barrier to fixation at $f = 0$ or $f = 1$. If genetic drift is added, random fluctuations in $f$ will lead to fixation at one of these absorbing boundaries. The change in the selection term $s(f)f(1-f)$ along the 'diagonal' $n_E = n_{EL} = n_{ET}$ is shown in Fig. S3B.

**Mutualism with a fixed point.** We first take a closer look at cases in which $s(f)$ has a mutualistic M region, where both amino acid concentrations are below the cross-over concentration required for maximum growth, i.e. $n_{EL} + n_{ET} \leq 2 + \kappa - \rho$, and $n_{EL}, n_{ET} < 1 + \kappa$. This mutualistic regime is characterized by a uniquely determined fixed point $f^*$ for which $s(f^*) = 0$:

$$f^* = \frac{1}{2} + \frac{1}{2} \frac{\kappa + 2}{2\rho(1+\kappa) + \kappa(n_{EL} + n_{ET})} (n_{EL} - n_{ET}) \qquad (36)$$

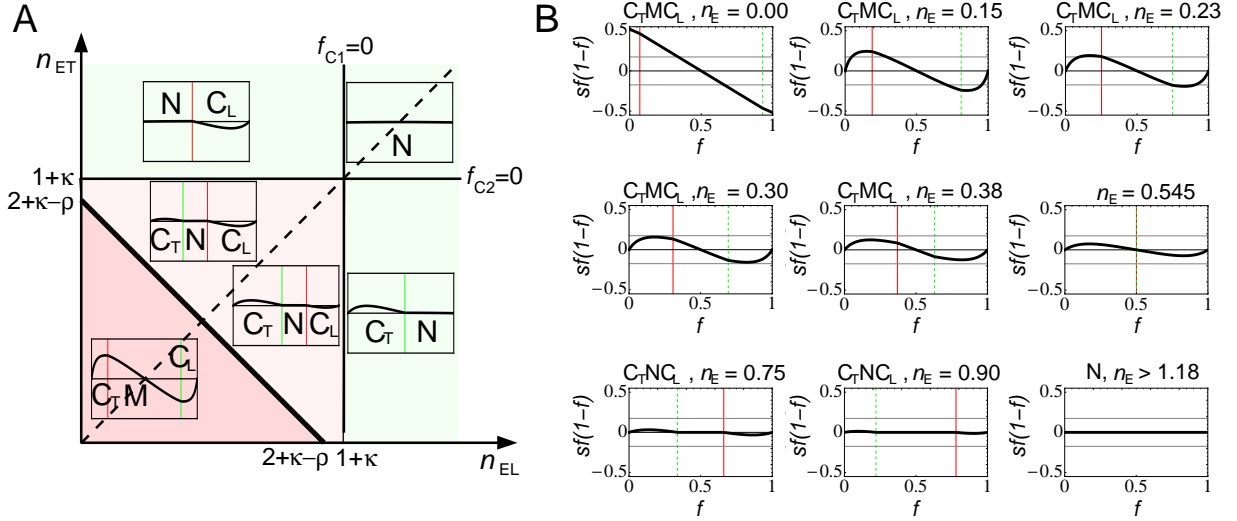

Figure S3: 'Phases' of mutualistic selection for yeast parameters as listed in Table S2. **A)** Deterministic 'phase diagram' in the plane of external amino acid concentrations $(n_{EL}, n_{ET})$. The insets show the selection term $s(f)f(1-f)$ in Equation (27) on a vertical scale of $-0.5$ to $0.5$ as a function of the fraction $f$ of leucine-requiring cells in the interval $0 \leq f \leq 1$. A positive selection term, $s(f)f(1-f) > 0$, acts to increase $f$, while negative values drive towards smaller $f$. The selective force vanishes for a selection term of zero, $s(f)f(1-f) = 0$. The selection term can be in one of five possible 'phases' designated as $C_T M C_L$, $C_T N C_L$, $C_T N$, $NC_L$, and $N$. In the phase names, M denotes the mutualistic regime in which both strains are amino-acid-limited, $n_L, n_T < 1$, so that they benefit from their mutualistic interaction. N represents the neutral regime with non-limiting amino acid concentrations $n_L, n_T > 1$ and therefore no mutualism. $C_T$ denotes a commensal regime in which the leucine requiring strain is saturated, $n_L > 1$, while the tryptophan-requiring strain is still limited, $n_T < 0$. $C_L$ describes the inverse situation with $n_L < 1$, $n_T > 1$. The vertical red and green lines in the insets mark the transition frequencies $f_{C1}$ and $f_{C2}$ between the different regimes. The red shades phases $C_T M C_L$ and $C_T N C_L$ represent mutualistic phases, in which selection drives the system towards a stable frequency $f^* \in (0, 1)$ within the M regime (dark red) or towards a neutral N region in the interior of the interval $f \in (0, 1)$ (light red). In the green shaded, non-mutualistic phases $C_T N$, $N$, and $NC_L$, the system has no barrier to fixation at $f = 0$ and/or $f = 1$. **B)** Selection term $s(f)f(1-f)$ along the diagonal $n_{EL} = n_{ET} \equiv n_E$. The vertical red and green lines in the insets mark the transition frequencies $f_{C1}$ and $f_{C2}$ as in (A). The black horizontal lines at $\pm s_c$, with $s_c = D_g^2/D_s = 0.18$, become comparable to the selection term around $n_E = 0.25$.

Because of the symmetry of the mutualistic interaction parameters ($\kappa = \kappa_L = \kappa_T$, $\rho = \rho_L = \rho_T$), the stationary allele frequency $f^*$ equals $1/2$ unless driven away from it by unequal external nutrient concentrations $n_{EL} \neq n_{ET}$. Vanishing selection $s(f) = 0$ in the mutualistic regime means that the amino acid concentrations $n_L(f^*) = n_T(f^*)$ are equal, see Equation (32):

$$n_L(f^*) = n_T(f^*) = \frac{n_{EL} + n_{ET} + \rho}{2 + \kappa}. \qquad (37)$$

As expected, the amino acid concentrations increase with the external concentrations $n_{EL}$, $n_{ET}$, and the cellular secretion rate $\rho$, but decrease with the uptake rate $\kappa$.

**Obligate mutualism.** In the obligate case, $n_{EL} = n_{ET} = 0$, the steady-state allele frequency is $f^* = 1/2$ due to the symmetry, as observed experimentally. The steady-state amino acid concentrations are $n_L(f^*) = n_T(f^*) = \rho/(2 + \kappa)$, so that the expansion velocity according to Equation (24) is $v \, n_L(f^*) = v \, n_T(f^*) = v \, \rho/(2 + \kappa)$. In our experiment, obligate mutualists expand with a velocity $v/2$ that is half the maximum velocity, see Figure 4F in the main text. Thus, we have

$$\rho = 1 + \frac{\kappa}{2}. \tag{38}$$

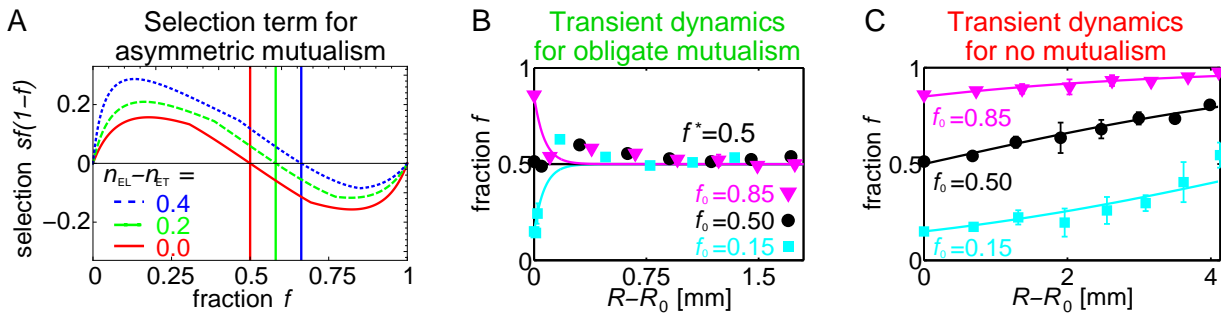

Figure S4: **A)** The frequency-dependent selection term $s(f)f(1-f)$ of Equation (27) for different amino acid asymmetries $n_{EL} - n_{ET}$, with fixed total amino acid amount $n_{EL} + n_{ET} = 0.6$. For increasing asymmetry $n_{EL} - n_{ET}$, fixation of leucine consumers at $f = 1$ becomes more likely, because (i) the stable fraction $f^*$ (indicated by vertical lines) of leucine consumers moves closer to $f = 1$, and (ii) the barrier towards fixation at $f = 1$ (the absolute value of the minimum of the selection term) decreases. **B,C)** Dynamics of the colony boundary fraction $f$, as function of radial distance $R - R_0$ from the homeland of radius $R_0$, for colonies with different inoculation fractions $f_0$. **B)** For obligate mutualism, the the boundary fraction reaches the steady-state fraction $f^* = 0.5$ within a 0.5mm radial expansion from the homeland with weak transient oscillations. **C)** For no mutualism, the boundary fraction increases during expansion because blue Leu$^-$ Trp$^{FBR}$ cells are 2% more fit than yellow Leu$^{FBR}$ Trp$^-$ cells. Data points in (B,C) are from flow cytometry of cells harvested from colony boundaries. Solid lines are solutions to selection dynamics with selection coefficient $s(f)$ of Equation (28) for obligate and $s = 0.02$ for no mutualism.

We use this condition to determine the value of the rescaled secretion rate $\rho$ from the value of $\kappa$. Equation (38) shows that for our mutualistic strains $\rho > 1$, so that the conditions $n_{EL}, n_{ET} > 1 - \rho$ in Equations (34) and (35) are always satisfied. In consequence, only the phases $C_T M C_L$, $C_T N C_L$, $C_T N$, $N C_L$, and $N$ are relevant for our system, as shown in Fig. S3.

**Absorbing boundaries for obligate mutualism.** In the obligate case $n_{EL} = n_{ET} = 0$, the selection coefficient in the mutualistic M regime in Equation (32),

$$s(f) = -\frac{1}{2} \frac{(1 + \kappa)(f - \frac{1}{2})}{(2 + \kappa)f(1 - f)}, \tag{39}$$

diverges when the fraction $f$ of leucine-requiring cells approaches 0 or 1. The reason is, for example in the case $f \to 0$, that the growth velocity $V_L$ of the rare Leu$^-$ Trp$^{FBR}$ cells approaches a constant, while the velocity $V_T$ of the abundant Leu$^{FBR}$ Trp$^-$ cells vanishes linearly with $f$. Similarly, the average population velocity $fV_L + (1 - f)V_T$ approaches zero linearly with $f$. Thus, as $f \to 0$, population growth ceases and thereby prevents the system from reaching the boundary $f = 0$ within finite time. A population with $f > 0$ will therefore never fix at $f = 0$. Note that the selection term $s(f)f(1 - f)$ in Equation (27) remains finite for all $f \in [0, 1]$.

**Asymmetric mutualism.** If e.g. more leucine than tryptophan is supplied in the medium, $n_{EL} - n_{ET} > 0$, mutualism is asymmetric, favoring leucine consumers, see Fig. S4. Compared to symmetric mutualism with a stable fraction $f^* = 0.5$, the stable fraction is closer to $f = 1$, and the barrier towards fixation at $f = 1$ (the maximum value of the absolute selection term in the interval $[f^*, 1]$) is smaller. Both effects make fixation of leucine consumers at $f = 1$ more likely. Likewise, tryptophan consumers are more likely to fix if there is more tryptophan than leucine in the medium.

**Mutualism with a neutral region.** The phase diagram in Fig. S3 exhibits an unusual phase $C_T N C_L$ in which the selection coefficient $s(f) = 0$ for a finite region in the interior of the interval $(0, 1)$. In this phase, selection in the commensal $C_T$ and $C_L$ regimes drives the system towards the neutral N regime, but dynamics within this regime is neutral. The existence of the neutral N regime is due to the plateaus in the velocity curves in Fig. S2: If both amino acid concentrations are $n_L$, $n_T \geq 1$, the precise concentration values do not matter since the colony always grows at maximum velocity.

**No mutualism.** For $n_{EL}$, $n_{ET} > 1 + \kappa$, the system does not have a barrier towards fixation at one or both of the absorbing frequency boundaries $f = 0$, $= 1$ (green regions in Fig. S3). In the $NC_L$ and $C_TN$ phases, selection acts towards the neutral N regime in which $s = 0$. Since the neutral regions connect to $f = 0$ and/or $f = 1$, number fluctuations can push the system towards an absorbing boundary, thus leading to demixing. In the N phase, genetic drift is the only acting force, leading to eventual fixation at $f = 0$ or $f = 1$.

### S3.2 The effect of small fitness differences

So far we have neglected the small fitness advantage $s_d = 0.02$ of the $Leu^- Trp^{FBR}$ strain over the $Leu^{FBR} Trp^-$ strain (Section S1.2). For low external amino acid concentrations, mutualism is a strong selective force with selection coefficients large compared to $s_d$, see Fig. S3, and it is justified to neglect this fitness difference. Experimentally, this approximation results in the observed symmetric steady-state cell fraction $f^* = 0.5$ 'despite' the fitness difference. However, for higher amino acid concentrations, the fitness difference becomes relevant as mutualistic selection becomes smaller.

**No mutualism.** For no mutualism, the fitness difference is the only selective force so that the dynamics of the fraction $f$ of leucine-requiring cells becomes

$$\frac{\partial f}{\partial \tau} = s_d\, f(1-f). \tag{40}$$

Thus, the frequency changes from the start frequency $f_0$ as

$$f(\tau) = \frac{f_0\, e^{s_d \tau}}{1 + f_0(e^{s_d \tau} - 1)} \tag{41}$$

and approaches fixation of $Leu^- Trp^{FBR}$, i.e. $f = 1$, for long times. This equation has been used to predict the cell fraction dynamics in Figure 3B in the main text and Fig. S4C.

**Weak mutualism.** The fitness difference is also important for weak mutualism, in particular when the selection coefficient has neutral regimes, as in the N, $C_TN$, $NC_L$, and $C_TNC_L$ phases: It 'tilts' flat the neutral regimes towards $f = 1$. This results in selection towards $f_{C2}$ in the $C_TNC_L$ phase, increases the likelihood for fixation at $f = 1$ in the N and $C_TN$ phases, while it decreases the probability of fixation at $f = 0$ in the $NC_L$ phase. This bias has been taken into account in the theoretical phase diagrams shown in Figure 4B in the main text, and explains at least part of the asymmetries in the experimental phase diagram in Figure 4A,B in the main text.

### S3.3 Transient approach to steady-state growth

Fig. S4B,C show the dynamics of the boundary fraction $f$ of blue cells as function of the radial distance from the homeland, which in our spatial expansion is analogous to time. In the case of obligate mutualism, the experimental boundary fraction converges to the steady-state value $f^* = 0.5$, independent of the inoculation fraction $f_0$. The approach happens rapidly, within less than 1 mm distance from the homeland with a small overshoot towards higher $f$, a feature expected for cross-feeding interactions [25]. Our simple mutualism model, designed for steady-state colony expansion, fails to produce this overshoot (although it does capture the time scale of the approach to steady-state, see solid lines in (B)), as well as asymmetric time lags to initiate growth (colonies with $f_0 = 0.15$ are smaller since they take longer to start growing, see Section S1.3, and colonies with $f_0 = 0.01$ do not grow at all).

In contrast to obligate mutualism, selection for no mutualism is not frequency-dependent, and the colony boundary fraction $f$ depends on the start fraction $f_0$ (Fig. S4C). The fraction $f$ of blue cells increases during expansion because the blue tryptophan producers have a 2% fitness advantage over the yellow leucine producers under these conditions (Section S1.2).

## S3.4  Antagonism between selection and genetic drift

As shown in Section S3.1, the mutualistic selection coefficient selects for 'mixing', i.e. for cell fractions $f$ in the interior of the interval $(0, 1)$, as long as the amino acid concentrations are below the crossover concentrations $n_{\text{EL}}, n_{\text{ET}} = 1 + \kappa = 1.18$. However, experimentally local demixing into domains with $f = 0$ or $f = 1$ occurs for much lower concentrations $n_{\text{EL}}, n_{\text{ET}} \geq 0.25$, see Figure 4 in the main text. This is due to the genetic drift term in Equation (27). In order to compare the strengths of mutualism and genetic drift, we non-dimensionalized Equation (27) by measuring distance in units of $D_{\text{s}}/D_{\text{g}}$ and time in units of $s_{\text{c}} \equiv D_{\text{s}}/D_{\text{g}}^2$ to obtain [22, 24]

$$\frac{\partial f}{\partial \widetilde{\tau}} = \frac{\partial^2 f}{\partial \widetilde{x}^2} + \frac{s(f)}{s_{\text{c}}} f(1-f) + \sqrt{f(1-f)} \, \Gamma(\widetilde{x}, \widetilde{\tau}). \tag{42}$$

with the dimensionless distance $\widetilde{x} = x/\left(D_{\text{s}}/D_{\text{g}}\right)$ and time $\widetilde{\tau} = t/\left(D_{\text{s}}/D_{\text{g}}^2\right)$ As shown in Ref. [24], the strength of mutualism and genetic drift become comparable if mutualistic selection becomes comparable to the critical selection coefficient

$$s_{\text{c}} \equiv \frac{D_{\text{g}}^2}{D_{\text{s}}} \tag{43}$$

that describes the strength of local demixing due to genetic drift. The 'mutualistic barrier' to demixing, i.e. fixation at $f = 0$ or $f = 1$ when the system is close to the stable fraction $f^*$, can be estimated as the maximum of the selection term $|s(f)f(1-f)|$, see Fig. S3B. One can see in Fig. S3B and in Figure 4D in the main text, that for symmetric mutualism with $n_{\text{E}} \equiv n_{\text{EL}} = n_{\text{ET}}$ this barrier becomes comparable to $s_{\text{c}}$ at about $n_{\text{E}} = 0.25$. Thus, we expect that mutualism is dominant for $n_{\text{E}} < 0.25$, leading to expansion in a mixed pattern, while genetic drift dominates for $n_{\text{E}} > 0.25$, leading to demixed expansion. This is indeed found experimentally, see Figure 4 in the main text.

For asymmetric mutualism, local fixation of the favored strain becomes more likely because the stable cell fraction $f^*$ moves closer to a fixation boundary $f = 0$ or $f = 1$, and because the 'mutualistic barrier' to fixation at that boundary becomes smaller, see Fig. S4. A more sophisticated analysis of the critical selection coefficient $s_{\text{c}}$ for asymmetric mutualism is the topic of a forthcoming theoretical paper [26].

## S4  Model for patches of obligate mutualists

In colonies of obligate mutualists, the two mutualistic strains form a pattern of blue and yellow 'patches', see Figure 2 in the main text. Because of the effective symmetry of our mutualistic interaction, yellow and blue patches have the same width, see Fig. S5A. After an initial transient of 1-4 days, the patch pattern becomes stationary, i.e. although patches may form and disappear, the overall pattern and patch widths remain statistically the same during the expansion.

In this section, we derive an approximation for the characteristic width of the patches during steady-state growth by considering the amino acid dynamics. Basically, the width of a patch of leucine-consuming cells is limited by how far leucine can diffuse into it from a neighboring leucine-producing patch before cellular uptake makes its concentration too low for cells in the patch interior to grow well. Although the patches are probably not completely demixed, here we first assume, for simplicity, that a yellow patch consists entirely of yellow Leu$^{\text{FBR}}$ Trp$^-$ cells, and a blue patch consists entirely of blue Leu$^-$ Trp$^{\text{FBR}}$ cells. We denote the average width of a blue and yellow patches parallel to the front as $L_{\text{L}}$ and $L_{\text{T}}$, respectively, see Fig. S5B.

**Diffusion with sources and sinks.** We first consider leucine, whose dynamics in the active layer of a colony is described by Equation (14). In a leucine 'production patch' of Leu$^{\text{FBR}}$ Trp$^-$ cells, we have $f = 0$ and thus

$$\text{Production patch:} \quad \frac{1}{d}\frac{\partial}{\partial t} n_{\text{L}} = \rho - (n_{\text{L}} - n_{\text{EL}}) + \frac{D_{\text{a}}}{d} \frac{\partial^2}{\partial x^2} n_{\text{L}}. \tag{44}$$

Similarly, the leucine dynamics in a leucine 'consumption patch' with $f = 1$ is given by

$$\text{Consumption patch:} \quad \frac{1}{d}\frac{\partial}{\partial t} n_{\text{L}} = -\kappa n_{\text{L}} - (n_{\text{L}} - n_{\text{EL}}) + \frac{D_{\text{a}}}{d} \frac{\partial^2}{\partial x^2} n_{\text{L}}. \tag{45}$$

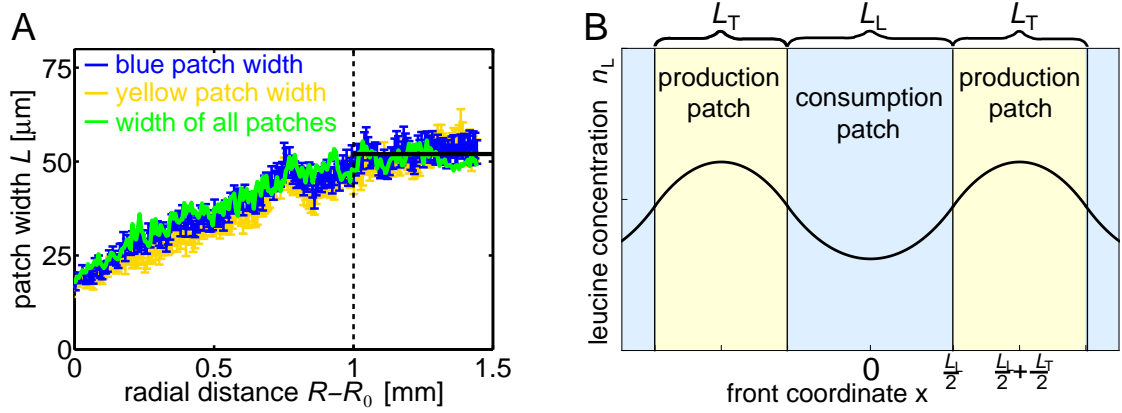

Figure S5: Determination of patch widths along the front coordinate $x$. **A)** Average patch widths (parallel to the front) as function of the radial distance $R - R_0$ from the homeland with radius $R_0$, determined via different methods for the same colony of obligate mutualists. Yellow patch width (yellow data) and blue patch width (blue data) are the average angular local maximum-to-minimum and minimum-to-maximum distances, respectively, of the derivative of the yellow fluorescence intensity. The width of all patches (green data) is reproduced from Figure 2D in the main text. Since it is is calculated as the circumference divided by number of local maxima of the yellow fluorescence intensity, it does not equal the average of the yellow and blue patch width. All patch widths are identical within error bars and saturate at $L = 52\,\mu$m (black horizontal line) for radial distances larger than 1mm. **B)** The amino acid leucine is secreted by Leu$^{\text{FBR}}$ Trp$^-$ in yellow production patches of width $L_{\text{T}}$, and consumed by Leu$^-$ Trp$^{\text{FBR}}$ cells in blue consumption patches of size $L_{\text{L}}$. The leucine concentration $n_{\text{L}}$ thus 'piles up' in the production patches, and decreases in the consumption patches. The width of the consumption patch is limited by this decrease, as explained in the text.

Here, the uptake rate in the leucine-limited regime ($n_{\text{L}} < 1$) has been used; for higher amino acid concentrations, the sink term would equal $-\kappa_{\text{L}}$.

As shown in Fig. S5B, a consumption patch is surrounded on both sides by production patches. Since we assume all consumption and production patches to have the same widths $L_{\text{L}}$ and $L_{\text{T}}$, respectively, the symmetry of the problem dictates that, in steady state, there is no amino acid current through the middle of the patches. Using the coordinate system shown in Fig. S5B, this leads to the boundary conditions

$$\left.\frac{\partial}{\partial x} n_{\text{L}}\right|_{x=0} = 0 \qquad \text{and} \qquad \left.\frac{\partial}{\partial x} n_{\text{L}}\right|_{x=\pm\frac{L_{\text{L}}+L_{\text{T}}}{2}} = 0. \tag{46}$$

In addition, the amino acid concentration and current have to be continuous at the patch boundaries:

$$n_{\text{L}} \text{ and } \frac{\partial}{\partial x} n_{\text{L}} \quad \text{continuous at } x = \pm\frac{L_{\text{L}}}{2} \tag{47}$$

**Solution to the diffusion problem.** Since we are interested in describing the pattern of the mutualistic strains during steady-state growth, we solve the diffusion equations (44) and (45) for the stationary state in which the amino acid profiles do not change over time. The diffusion problem (44,45) with the boundary conditions (46,47) can be solved analytically in terms of hyperbolic functions. The minimal and maximal leucine concentrations within the consumption patch are attained in the middle $x = 0$ and at the boundary $x = L_{\text{L}}/2$ of the consumption patch, respectively. For obligate mutualism with $n_{\text{EL}} = 0$, these values can be obtained from the full analytic solution:

$$n_{\text{L,min}} = n_{\text{L}}(0) = \frac{\rho}{\Sigma} \sinh\left(\frac{L_{\text{T}}}{2\xi_{\text{P}}}\right) \tag{48}$$

$$n_{\text{L,max}} = n_{\text{L}}(L_{\text{L}}/2) = \frac{\rho}{\Sigma} \sinh\left(\frac{L_{\text{T}}}{2\xi_{\text{P}}}\right) \cosh\left(\frac{L_{\text{L}}}{2l_{\text{a}}}\right) \tag{49}$$

where

$$\xi_{\text{p}} = \sqrt{\frac{D_{\text{a}}}{d}} \qquad \text{and} \qquad l_{\text{a}} = \sqrt{\frac{D_{\text{a}}/d}{\kappa + 1}} \qquad (50)$$

are the length scales characterizing the 'piling up' of leucine in the production patch and the decay of the leucine concentration in the consumption patch, respectively. We also used the abbreviation

$$\Sigma = \cosh\left(\frac{L_{\text{L}}}{2l_{\text{a}}}\right) \sinh\left(\frac{L_{\text{T}}}{2\xi_{\text{p}}}\right) + \frac{\xi_{\text{p}}}{l_{\text{a}}} \sinh\left(\frac{L_{\text{L}}}{2l_{\text{a}}}\right) \cosh\left(\frac{L_{\text{T}}}{2\xi_{\text{p}}}\right). \qquad (51)$$

**Velocity gradients within consumption patches.** In the amino acid-limited regime, the growth velocity is proportional to the concentration of the required amino acid, see Equations (1,2). This means that the growth velocity varies within a consumption patch: It is smallest in the center and largest at the boundary. If cells grew only in the expansion direction, these velocity differences along the colony front would lead to non-uniform front-propagation and thus to an unstable, undulated colony front. However, experiments show that the obligate mutualists expand together with a smooth colony front, see Figure 2 in the main text. In addition, the velocity differences generate a selection force perpendicular to the expansion direction. In a leucine consumption patch, this perpendicular velocity gradient can be estimated by [6, 27]

$$V_{\perp \text{L}} = v_{\text{L}} \sqrt{n_{\text{L,max}}^2 - n_{\text{L,min}}^2}. \qquad (52)$$

There must be a force that counteracts the selective force generated by the velocity differences within a path.

**Patch boundary diffusion.** Colony growth of non-motile yeast cells proceeds due to cell division, which results in pushing and shuffling of cells. This leads to a diffusion-like process that occurs parallel to as well as perpendicular to the expansion direction. Experimentally, this diffusive mixing can be observed as the diffusion of sector boundaries with diffusion constant $D_{\text{s}}/\tau_g$, where $\tau_g$ is the generation time at the patch boundary [27, 28]. Since the growth velocity at the boundary equals $v = v_{\text{L}} n_{\text{L,max}}$, and since the growth rate at the boundary is $g = 1/\tau_g = v/b$ (Section S1.3), we can estimate the diffusion 'velocity' required to diffuse on a length scale of half a patch of width $L_{\text{L}}$ as

$$V_{\text{DL}} = \frac{4\,(D_{\text{s}}/\tau_g)}{L_{\text{L}}} = \frac{4 D_{\text{s}} \, v_{\text{L}} n_{\text{L,max}}}{b\, L_{\text{L}}} = \frac{2 l_{\text{b}} \, v_{\text{L}} n_{\text{L,max}}}{L_{\text{L}}}, \qquad (53)$$

where we have introduced the boundary diffusion length scale $l_{\text{b}} = 2 D_{\text{s}}/b$.

**Balance condition determines patch width.** This diffusive process can counteract the imbalances created by the velocity differences if both velocities are of the same order of magnitude,

$$V_{\perp \text{L}} = V_{\text{DL}}. \qquad (54)$$

An analogous equation holds for the width $L_{\text{T}}$ of tryptophan consumption patches. In the case of obligate mutualism, i.e. no external amino acids $n_{\text{EL}} = n_{\text{ET}} = 0$, the mutualistic interaction is fully symmetric so that $L_{\text{L}} = L_{\text{T}} = L$. In this case, Equation (54) yields

$$L = 2\, l_{\text{b}} \, \coth\left(\frac{L}{2 l_{\text{a}}}\right), \qquad (55)$$

which can be solved approximately for small patch widths to yield Eq. [5] in the main text,

$$L = 2\, \sqrt{l_{\text{a}}\, l_{\text{b}}}. \qquad (56)$$

## S5 Radial flux of amino acids into and out of a colony

In this section, we derive an approximate description for the diffusive flux of amino acids into and out of the actively growing boundary layer of a colony. We will show that, on the experimental timescales of many hours to days, the amino acid diffusion dynamics can be approximately described by a chemostat-like process as used in Eq. [1] in the main text, and we will derive an expression for the effective influx / outflux rate $d$ in terms of the colony growth parameters.

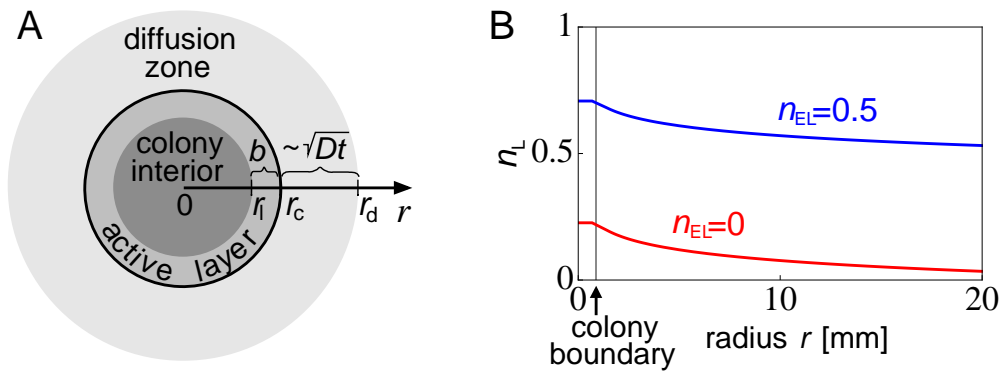

Figure S6: **A)** Geometry to calculate amino acid diffusion within and out of a colony. In a colony of radius $r_c$, only cells in an *active layer* of size $b = r_c - r_l$ near the colony boundary grow, while cells in the *colony interior* $r \leq r_l$ do not grow due to depletion of essential nutrients (e.g. glucose). In the active layer, amino acids diffuse, and are produced and taken up by cells. Outside the colony, amino acids diffuse freely and create a *diffusion zone* of radial size $r_d - r_c \sim \sqrt{4D_a t}$. **B)** Radial amino acid concentration profile $n_L(r) = [\text{leu}]_r/[\text{leu}]_c$ for yeast parameters (Table S2) for external nutrient concentrations $n_{EL} = [\text{leu}]/[\text{leu}]_c = 0$ (red) and $n_{EL} = 0.5$ (blue).

## S5.1 Nutrient diffusion and colony growth

**Diffusion time scales.** The cells in a growing yeast colony obtain their nutrients, e.g. glucose and amino acids, from the agar medium on which they grow. Nutrients reach cells via diffusion with a diffusion constant $D_{\text{nut}}$ of a few mm$^2$/h [15, 29]. . This parameter value means that diffusion can sample the 5 mm height of the agar in the Petri dish within a few generations (generation time $\tau_g \approx 1.5$h). Thus, on experimental timescales of several days, we integrate out the diffusion perpendicular to the agar surface and only consider the two-dimensional diffusion of nutrients in the plane of the colony, see Fig. S6. In addition, diffusion is fast compared to the colony expansion velocity of $v \approx 20\mu$m/h: a colony will outrun nutrient diffusion only after about $D_{\text{nut}}/v^2 \approx 1$ year. We can therefore ignore colony expansion for nutrient dynamics on experimental timescales of several days.

**The active layer.** We first consider the diffusion and uptake of an essential nutrient that is not produced by the cells, such as glucose. After an initial transient of 1-2 days, in which the colony depletes the glucose beneath it, new glucose reaches the colony by radial diffusion towards the colony. This glucose allows only the cells in the *active layer* of size $b = 40\,\mu$m near the colony boundary to grow [11, 12]. Consistent with the chemostat-like nature of the frontier region, these cells in the colony boundary consume all the glucose, so that cells in the colony interior cannot grow. This geometry is shown in Fig. S6A.

## S5.2 Radial amino acid dynamics

We now consider the radial diffusion dynamics of leucine (tryptophan works analogously), which is produced or taken up by cells in the active layer of a colony consisting of one or both of the mutualistic strains. Here, we are interested in length scales larger than the angular width of patches or sectors. We therefore consider the cells in the active layer to be essentially well-mixed, so that they can be described by the $r$-independent concentrations $c_L$ and $c_T$ of Leu$^-$ Trp$^{\text{FBR}}$ and Leu$^{\text{FBR}}$ Trp$^-$ cells, respectively. The two-dimensional dynamics of the radial leucine concentration $[\text{leu}]_r \equiv [\text{leu}](r,t)$ in the active layer can then be described by the radially symmetric diffusion equation,

$$\text{for } r_l < r < r_c: \qquad \frac{\partial}{\partial t}[\text{leu}]_r = D_a \frac{1}{r} \frac{\partial}{\partial r}\left(r \frac{\partial}{\partial r}\right)[\text{leu}]_r - \frac{k_L}{[\text{leu}]_c} c_L\,[\text{leu}]_r + r_L c_T. \qquad (57)$$

where $r_c$ is the colony radius and $r_c - r_l = b$ is the size of the active layer. Outside of the colony, leucine diffuses freely,

$$\text{for } r \geq r_c: \quad \frac{1}{D_a}\frac{\partial}{\partial t}[\text{leu}]_r = \frac{1}{r}\frac{\partial}{\partial r}\left(r\frac{\partial}{\partial r}\right)[\text{leu}]_r. \tag{58}$$

This diffusion causes a *diffusion zone* around the colony, within which the leucine concentration differs from its concentration $n_{\text{EL}}$ in the agar medium far away from the colony (the concentration with which the medium was prepared). On the experimentally relevant times of 4-7 days, the size of the diffusion layer, $r_d - r_c \sim \sqrt{4D_a t}$, increases very slowly from 34mm to 44mm. For distances larger than this, the leucine concentration equals its value far away from the colony,

$$[\text{leu}]_r(r \geq r_d) = [\text{leu}]_E. \tag{59}$$

In a quasi-stationary approximation, we can solve the diffusion equations (57,58) for their stationary states with the boundary condition (59). In addition to the boundary condition (59), the amino acid concentration and current have to be continuous at the connection between the active layer and the diffusion zone:

$$[\text{leu}]_r, \ \frac{\partial[\text{leu}]_r}{\partial r} \ \text{ continuous at } \ r = r_c. \tag{60}$$

Furthermore, since cells in the colony interior are inactive, amino acids diffuse freely for $r < r_l$, with a zero-current condition in the colony center at $r = 0$ because of symmetry. This translates into a zero-current condition at the boundary between the active layer and the colony interior:

$$\frac{\partial[\text{leu}]_r}{\partial r}(r = r_l) = 0. \tag{61}$$

**Solution to the diffusion problem.** The diffusion problem (57,58) with the boundary conditions (59-61) can be solved analytically. The concentration within the active layer is a superposition of the modified Bessel functions of the first and second kind, while the concentration decays logarithmically in the diffusion zone. It is plotted for yeast parameters in Fig. S6B.

**Flux into and out of the active layer.** As can be seen in Fig. S6B, radial diffusion happens on length scales large compared to the radial size of the active layer $b = 40\,\mu$m. Thus, we can approximate the leucine concentration within the active layer, $n_L$, by the concentration at the colony boundary, $[\text{leu}] \equiv [\text{leu}]_r(r_c)$. The flux at the colony boundary can be calculated from the exact solution to the diffusion problem and has the simple form

$$j_L \equiv j_L(r_c) = D_a\frac{\partial[\text{leu}]_r}{\partial r}\bigg|_{r=r_c} = -\frac{D}{r_c \ln\left(\frac{r_d}{r_c}\right)}\left([\text{leu}] - [\text{leu}]_E\right). \tag{62}$$

Due to the boundary flux $j_L$, the number of leucine molecules in an area element $b\,\Delta x$ of the active layer changes in a time increment $\Delta t$ according to

$$[[\text{leu}](t + \Delta t) - [\text{leu}](t)]\, b\,\Delta x = j_L\,\Delta x\,\Delta t, \tag{63}$$

or

$$\frac{\partial}{\partial t}[\text{leu}] = \frac{j_L}{b} = -d\left([\text{leu}] - [\text{leu}]_E\right) \tag{64}$$

with the diffusive flux constant

$$d = \frac{D_a}{b\,r_c \ln\left(\frac{r_d}{r_c}\right)} = \frac{D_a}{b\,r_c \ln\left(1 + \frac{\sqrt{4D_a t}}{r_c}\right)}. \tag{65}$$

This diffusive flux is balanced by leucine secretion and uptake in the active layer, leading to an effective dynamic equation for the nutrient concentration $n_L$ in the active layer

$$\frac{\partial}{\partial t}[\text{leu}] = -k_L c_L\,[\text{leu}] + r_L c_T - d\left([\text{leu}] - [\text{leu}]_E\right). \tag{66}$$

A similar equation holds for tryptophan dynamics. With the amino acid diffusion constant $D_a = 3\,\text{mm}^2/\text{h}$, the size $b = 40\,\mu$m of the active layer, a colony radius of $r_c = 3 - 5\,\text{mm}$, and the size $\sqrt{4D_a t} \approx 34 - 44\,\text{mm}$ of the diffusion zone, the diffusive flux constant is of the order of $d \approx 5/\text{h}$.